\documentclass{ieeeaccess}
\usepackage{cite}
\usepackage{amsmath,amssymb,amsfonts}
\usepackage{algorithmic}
\usepackage{graphicx}
\usepackage[table,xcdraw]{xcolor}
\usepackage{capt-of,lipsum} 
\usepackage{textcomp}
\usepackage{subfigure}
\usepackage{caption,setspace}
\def\BibTeX{{\rm B\kern-.05em{\sc i\kern-.025em b}\kern-.08em
    T\kern-.1667em\lower.7ex\hbox{E}\kern-.125emX}}

\newcommand{\ie}{\textit{i}.\textit{e}.,\ }
\newcommand{\eg}{\textit{e}.\textit{g}.,\ }
\newcommand{\cf}{\textit{c}\textit{f}.\ }
    
\begin{document}
\history{Date of publication xxxx 00, 0000, date of current version xxxx 00, 0000.}
\doi{10.1109/ACCESS.2017.DOI}

\title{Real-time Tracking of Medical Devices: An Analysis of Multilateration and Fingerprinting Approaches}

\author{
    Bruno Rodrigues, Eder J. Scheid, Katharina O. E. M{\"u}ller, Julius Willems, Burkhard Stiller
}
\address{Communication Systems Group CSG, Department of Informatics IfI, University of Z{\"u}rich UZH\\
Binzm{\"u}hlestrasse 14, CH---8050 Z{\"u}rich, Switzerland\\
}

\markboth
{Rodrigues \headeretal: Preparation of Papers for IEEE TRANSACTIONS and JOURNALS}
{Rodrigues \headeretal: Preparation of Papers for IEEE TRANSACTIONS and JOURNALS}

\corresp{E-mail: [rodrigues$\mid$scheid$\mid$mueller$\mid$stiller]@ifi.uzh.ch, julius.willems@uzh.ch}

\begin{abstract}
Hospital infrastructures are always in evidence in periods of crisis, such as natural disasters or pandemic events, under stress. The recent COVID-19 pandemic exposed several inefficiencies in hospital systems over a relatively long period. Among these inefficiencies are human factors, such as how to manage staff during periods of high demand, and technical factors, including the management of Portable Medical Devices (PMD), such as mechanical ventilators, capnography monitors, infusion pumps, or pulse oximeters. These devices, which are vital for monitoring patients or performing different procedures, were found to have a high turnover during high-demand, resulting in inefficiencies and more pressure on medical teams.

Thus, the work PMD-Track evaluates in detail two popular indoor tracking approaches concerning their accuracy, placement of beacons, and economic impacts. The key novelty of PMD-Track relies on using smartphones provided to hospital employees, replacing typical stationary gateways spread across a hospital, functioning as mobile gateways with a front-end that assists staff in locating PMDs. As employees approach tagged PMDs, their smartphone automatically updates the location of spotted PMDs in real-time, providing room-level localization data with up to 83\% accuracy for fingerprinting and 35\% for multilateration. In addition, fingerprinting is 45\% cheaper than multilateration over the course of five years. Practical experiments were evaluated based on two locations in Zürich, Switzerland. 
\end{abstract}


\titlepgskip=-15pt

\maketitle

\section{Introduction} \label{chp:introduction}

The recent global pandemic exacerbated the challenge of hospital equipment management. To ensure a sufficient availability of medical equipment and to cope with related inefficiencies, hospitals typically acquire excess capacities, resulting in cost overheads and asset utilization rates below 50\%. At the same time, experts consider a utilization rate of 80\% feasible \cite{horblyuk2012out}. The recent period of scaling demands revealed various flaws in healthcare systems worldwide due to the unexpected rise in patient load, whose inefficiencies to respond effectively during times of crisis were pointed out~\cite{barach2020disruption,sharma2021responses}. For example, it was found  that mechanical ventilators and supervision equipment, such as ECG (Electrocardiogram), capnography monitors, infusion pumps, or pulse oximeters, were not adequately managed on hospital property and that, in the majority of cases, hospitals were understaffed \cite{barach2020disruption}.

\begin{figure}[ht]
  \centering
  \resizebox{\columnwidth}{!}{%
    \includegraphics[width=.75\linewidth,keepaspectratio]{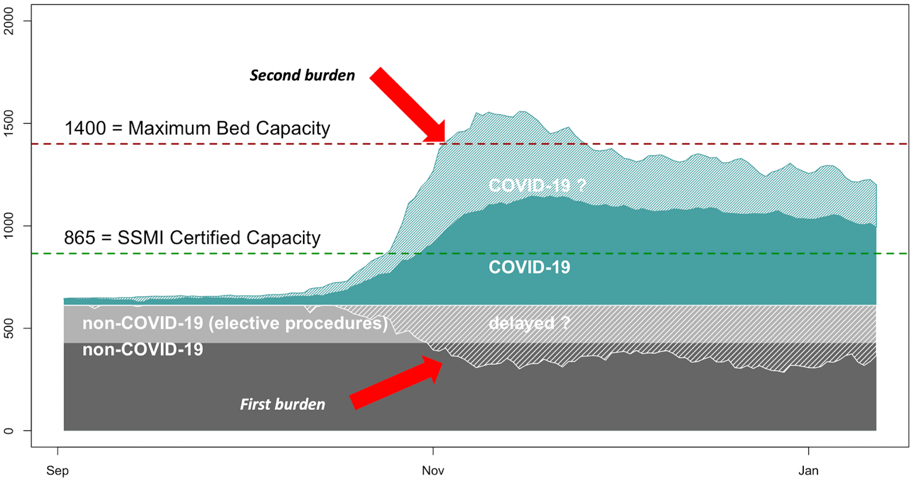}
  }
  \caption{Occupancy of intensive care units in Switzerland in 2020 \cite{van2021double}}
  \label{fig:hospital-burden}
\end{figure}

Figure \ref{fig:hospital-burden} presents a hospital occupancy for September 2020 to January 2021, in which the occupancy exceeds not only the capacity certified by the Swiss Society of Intensive Care Medicine (SSMI), but also the maximum capacity \cite{van2021double}. The period of great stress on healthcare infrastructures has triggered several considerations to operate more efficiently, such as managing available infrastructure and devices and the overwhelming stress on the medical staff that influences several management aspects~\cite{kuo2020survey}.  

During periods of high demand, such as health crises during the COVID pandemic or natural disasters such as earthquakes, Portable Medical Devices (PMD) for monitoring patients or performing different procedures were found to have a high turnover \cite{sharma2021responses}. Staff typically communicates about such devices orally, transferring the equipment's responsibility when their shift ends \cite{yoo2018real}. The lack of an automated and structured approach for locating portable equipment and the pressure the staff is typically exposed to during these periods makes their management ineffective. 

\subsection{Benefits and Drawbacks}

Tracking the location of PMDs automatically increases a hospital's efficiency in reacting to emergencies. The main \textbf{advantage} for using tracking solutions is that these show  potential for increasing operational efficiency to quickly react to emergencies by knowing about precise locations of  PMDs~\cite{barach2020disruption,sharma2021responses}. While hospitals do not always operate close to their capacity (often measured based on the number of occupied beds \cite{ravaghi2020models}), in exceptional cases, it is common for this capacity to be reached or even exceeded, and tracking solutions can have a significant positive impact in terms of saving lives. In addition, there are additional benefits, such as reducing the burden on medical staff by decreasing unnecessary communications and the time required to find equipment, improving inventory management, and potentially reducing overall expenses in the long-term once hospitals can better estimate equipment replacement and maintenance. 

While there are several benefits, there are also \textbf{drawbacks} that prevent such solutions from being deployed on a large scale in hospitals. The literature presents several solutions based on different fingerprinting and multilateration approaches, using technologies such as RFID (Radio-Frequency IDentification), UWB (Ultra Wide-band), WiFI, and Bluetooth Low Energy (BLE) for device tracking \cite{dardari2015indoor}. Approaches have also been proposed in the context of hospitals but have not seen a widespread deployment \cite{oztekin2010rfid,carrasco2010real,youn2007wlan,przybylo2014smarter}. For instance, \cite{oztekin2010rfid} uses RFID tags attached to assets, herein termed PMDs, and RFID readers scattered through the hospital's infrastructure to read the position of tags. In a similar direction, \cite{youn2007wlan} proposed a WiFi tracking solution using multilateration to find the position of wheelchairs in real-time. While RFID shows the drawback of requiring several RFID readers, a WiFi-based approach offers a higher range, but suffers from a lack of accuracy.

Since hospitals typically do not operate at full capacity, and the cost of deploying and operating these solutions is relatively high, the need for precise tracking solutions is often not perceived as necessary by hospital managers. Thus, instead of improving how existing PMDs are managed, a typical approach is often to acquire more PMDs. However, the need to improve how PMDs are managed became apparent during the recent pandemic. Several hospital infrastructures were overwhelmed, including staff communications under stressful situations ~\cite{barach2020disruption,sharma2021responses}, thus, resulting in ``missing" PMDs, which only had been misplaced. In addition, the effects of increased signal emission still require further study to ensure that its operation does not affect patients' health in critical states. For example, \cite{christe2010analysis} has run a descriptive study on the impact of RFID asset-tracking in healthcare, but the area still requires practical studies and larger analysis.

\subsection{Overview and Contributions}

PMD-Track's initial design was first presented in \cite{rodrigues2022pmd}, and a preliminary evaluation was published in \cite{pmdExperience2023}. Nevertheless, this paper presents the full architecture of PMD-Track, and a very detailed real-world evaluation of its two prototyped tracking approaches, \textit{(i)}~fingerprinting and \textit{(ii)}~multilateration. Furthermore, this paper includes an economic analysis considering the number of devices necessary to operate a hospital effectively. Other and previous work of the authors covered different aspects and tracking technologies, such as \textit{(a)} passive Bluetooth and WiFi  tracking~\cite{rodrigues2021bluepil,ribeiro2021asimov}, \textit{(b)} the combined use of RFID and cameras tracking~\cite{RCFSKAS22}, and \textit{(c)} the correlation of several tracking sources using temporal and spatial dimensions to improve precision~\cite{RSWTMS22}. Although these approaches use passive tracking in the context of event marketing analysis, instead of the hospital context, the algorithms and techniques developed contributed indirectly to the PMD-Track's design.

The main challenge that PMD-Track approach faced was the cost/benefit optimization of existing asset-tracking approaches used in hospitals. In this regard, PMD-Track leverages two fundamental pillars: tracking operation based on smartphones and static BLE tags components that simplify the operation in contrast to existing indoor tracking solutions, combined with an intuitive visualization and detailed analytics on the usage of PMDs. The location of tagged PMDs can be automatically updated whenever a staff member passes by or comes within the range of a tagged PMD. Hospitals typically equip their staff with smartphones to facilitate internal communications and provide easy access to hospital services. This is the key aspect this proposal considers: smartphones can replace expensive gateways scattered throughout the hospital's infrastructure by acting as mobile gateways. 
This paper's \textbf{contributions} are summarized as follows:

\begin{itemize}
    \item Design and prototyping of a gateway-less tracking approach, replacing gateways typically used in real-time tracking solutions with mobile devices used by hospital employees.
    \item Comparison of room-level accuracy, training size, and beacon placement of multilateration and fingerprinting tracking approaches in a real-world evaluation.
    \item Presenting an extensive analysis of the proposed approach regarding the accuracy, economics, and impacts on security and privacy.
\end{itemize}

\subsection{Organization}

The remainder of this paper is organized as follows. Section~\ref{sec:fundamentals} overviews fundamentals. While Section~\ref{sec:design} describes the rationale and design, Section~\ref{sec:evaluation} details the evaluation in essential dimensions. Finally, Section~\ref{sec:conclusions} summarizes the work and outlines future steps.

\section{Fundamentals}
\label{sec:fundamentals}

While Subsection \ref{subsection:background} provides insights into major concepts required for this work, especially Bluetooth Low Energy (BLE) and the comparison of tracking approaches \ie multilateration and fingerprinting, Subsection \ref{subsection:related-work} surveys related work in indoor tracking and within hospitals, as well as respective visualization and analytics.

\subsection{Background}
\label{subsection:background}

The major and underlying concepts include Bluetooth Low Energy (BLE) used in tags (\cf~ Subsection~\ref{subsec:ble}) and tracking approaches of multilateration (\cf Subsection~\ref{subsec:multilateration}) as well as fingerprinting (\cf Subsection~\ref{subsec:fingerprinting}).

\subsubsection{Bluetooth Low Energy (BLE)}
\label{subsec:ble}
BLE is a widely adopted wireless technology for personal area networking with a range of up to 100~m in line-of-sight situations. It operates on 40 channels in the unlicensed 2.4~GHz ISM frequency band. 
BLE is made for low-power data transmission at up to 2~Mbit/s, making it one of the most popular technologies in Internet of Things (IoT) applications \cite{learn-about-bluetooth}. Recent estimates indicate that by 2026, 7 billion BLE-enabled devices will be shipped each year, a common communication protocol embedded in IoT devices (\eg smart home, industrial IoT) \cite{stojkoska2017review,komninos2014survey}.

\subsubsection{Multilateration}
\label{subsec:multilateration}
Is an approach to geometrically estimate an object's position in space through distance measures to at least three points (\ie trilateration is the minimal case). This corresponds to solving the following non-linear system, with $(x_i,y_i,z_i)$ being the position of the $i$-th point, $(x,y,z)$ as the position of the object, and $d_i$ as the distance of the object to the $i$-th point. In a two-dimensional (planar) space, this leads to solving the following system in two variables:
\begin{equation} \label{eq:trilateration-planar}
    \begin{split}
        (x - x_1)^2 + (y - y_1)^2 &= d_1^2 \\
        (x - x_2)^2 + (y - y_2)^2 &= d_2^2 \\
        (x - x_3)^2 + (y - y_3)^2 &= d_3^2
    \end{split}
\end{equation}

\begin{figure}[ht]
    \centering
    \includegraphics[width=.9\linewidth]{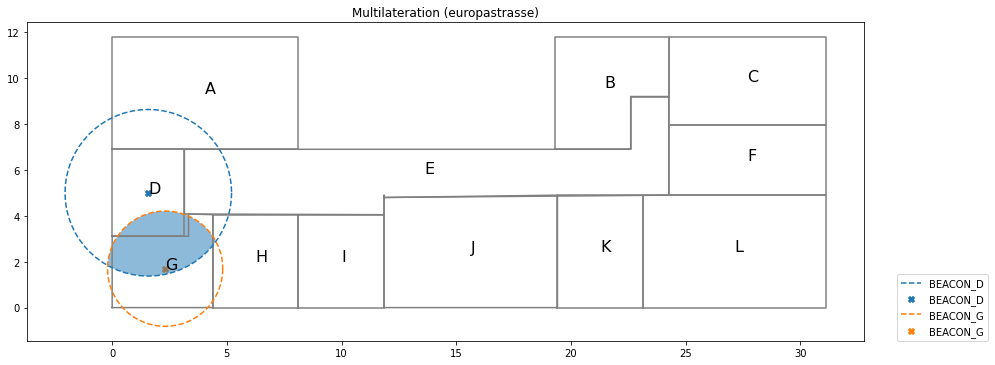}
    \caption{Multilateration example in the evaluation scenario detecting a device in the blue area.}
    \label{fig:multi_case_3}
\end{figure}

Solving equation \ref{eq:trilateration-planar} determines $x$ and $y$ coordinates of point~P. This is observed in Figure \ref{fig:multi_case_3}, in which an item attached to a BLE beacon's position is predicted within the blue area. The position can be calculated using the location of at least three stationary beacons to reach the solution analogously for any higher-order multilateration problems, \ie four or more beacons. In practice, distance measures are often imperfect, once signal measurements often present slight variances, and the calculation of a solution for using a non-linear optimization leads to better results \cite{dardari2015indoor}.

\subsubsection{Fingerprinting}
\label{subsec:fingerprinting}
Is a method based on pattern recognition and consists of (a) an offline or training phase and (b) an online or operational phase. In (a), a site survey is conducted where measurements are collected in areas of interest. These measurements are called fingerprints and are ideal as uniquely as possible for each area. A typical example of fingerprinting is WiFi fingerprinting, where the Received Signal Strength Indicator (RSSI) of different Access Points (AP) is measured at specific points. In (b), a user can record the fingerprint at its current location and query it against the fingerprint database. 

\begin{figure}[ht]
  \centering
  \includegraphics[width=.4\linewidth,keepaspectratio]{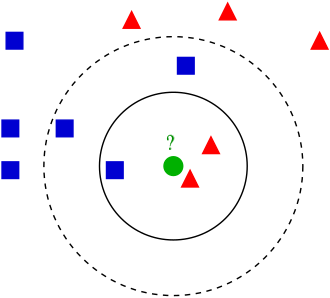}
  \caption{kNN example}
  \label{fig:knn_example}
\end{figure}

The k-nearest neighbors (kNN) classifier is used for the fingerprinting model, and a modified implementation of a multilateration technique for approximating the room based on geometric calculations \cite{ge2016optimization, bi2018adaptive, liang2012fingerprint}. In the example shown in Figure \ref{fig:knn_example}, the unseen green data point is classified as a red triangle for \(k=3\) or a blue square for \(k=5\). kNN uses distance-based metrics to determine neighboring data points. In the case of the collected RSSI values, the data set is split into a disjoint training and test data set with the dimensions \(cn \times{k+1}\) and \((1-c)n \times{k+1}\) for a train-test-split coefficient \(c\) (\ie $c=0.2$).

\[class(q) = majority(k\_min(\parallel q-v_i\parallel)) \forall v_i \in train\]
The training phase consists of storing the k-dimensional training samples and their associated class labels. In the classification phase of an unlabeled vector \(q\), its k nearest neighbors are calculated from the training data set based on the chosen distance metric. Eventually, the class label of \(q\) is determined based on most of its nearest neighbor's class labels. 

\subsection{Related Work}
\label{subsection:related-work}

As indoor location services gain popularity due to the absence of Global Positioning System (GPS) signal indoor, BLE and RFID (Radio Frequency Identification) technologies emerge as an alternative (among others such as WiFi, Ultra-WideBand (UWB), and Visible Light Communication (VLC). Among these technologies, BLE provides the most attractive trade-off between cost and accuracy at the cost of having a relatively low range.

\subsubsection{WiFi and Bluetooth-based Approaches}

\cite{rodrigues2022pmd} presented previous work on the initial architecture of PMD-Track. Although presenting the idea of a gateway-less architecture based on BLE beacons and BLE-enabled mobile phones, the paper lacked an in-depth analysis that stems from a real-life implementation and experimentation, as presented in this article. In this regard, this article presents a detailed architecture and evaluation comparing fingerprinting and multilateration approaches. 

\cite{rodrigues2021bluepil} developed a passive approach to track devices emitting Bluetooth packets. Positive points of this work are characterized by the architecture of streaming data (from multiple sensors) to a sink where the location data is processed. However, a passive approach is imprecise because it requires devices to emit packets to capture them in the environment. Thus, such a solution would not be feasible to track assets within a hospital since tags can be paired with nearby sinks, eliminating the need for a passive approach.

\cite{yoo2018real} developed a system using active BLE beacons attached to medical devices and the hospital's WiFi access points to relay data. Then, the authors relied on signals actively emitted by beacons to calculate their position using a multilateration approach. The authors identified two major issues: poor precision due to interference from other radio waves and relatively high battery consumption considering active beacons. One disadvantage of the multilateration approach is that it requires constant probing of devices. In the authors' case, the devices report their position directly to the APs. Further, the fact that the beacons connect to APs means that they present an onboard network interface, which impacts the power consumption and the price of each tag.

\cite{canton2017bluetooth} proposes a real-time indoor positioning system based on BLE using frequency diversity, trilateration, and Kalman Filter (KF). The tag position is calculated in the approach based on the trilateration of 3 RSSI sniffer devices and the tag sending RSSI values; KF is used to smooth the position calculations. One shortcoming of the approach is that it relies on four beacons per room, and each receiver (\ie sniffer device) costs around 120 Euros. This makes the approach not cost-effective to be employed in the PMD-Track as its goal is to reduce costs.

\cite{adjei2020developing} proposes a real-time tracking solution based on Arduino to track Bluetooth devices at a maximum range of 10 meters. The Arduino board also contains a GSM antenna that periodically transmits the collected data to a user's smartphone. The authors' approach has two problems that PMD-Track solves: the possible use of several static sinks (Arduino devices) to collect data from BLE tags, and the limited mobility combined with the short-range makes the solution ineffective for tracking objects in the hospital scenario.

\cite{youn2007wlan} presents real-time asset tracking based on WiFi tags and the existing WiFi infrastructure available in hospitals. The authors employed fingerprinting based on RSSI signals from six beacon APs covering an area of 450 square meters \ie entire floor of a medium-sized hospital floor (60 rooms). Thus, positioning data was collected during the training stage, and real-time measurements were compared and approximated with training values to determine their position. However, WiFi-based RSSI is highly unreliable due to the number of running devices and connections, an aspect that the authors pointed out during their evaluation.

\subsubsection{RFID-based Approaches}

\cite{hakim2006passive} presents a passive RFID tracking system for hospitals. While passive tags present the benefits of being relatively small and not requiring batteries, they also are relatively less reliable and accurate than active tracking solutions. The authors tested the proposed system in the university's biomedical department. Although a financial analysis was included, the publication did not disclose an evaluation in terms of the accuracy and precision of the proposed approach. In this regard, passive RFID tracking systems are known for their poor accuracy, given that tags rely on the active signal of nearby readers and other factors related to the interference of signals \cite{dardari2015indoor}. This fact can also be observed in CCount \cite{RCFSKAS22}, in which a combination of passive RFID tags and cameras were used to track people's movement at indoor events and gauge interest in products and merchandise.

\cite{bisio2016new} developed a crowdsourced asset-tracking solution for the construction industry based on integrating RFID and BLE technology. Assets (\eg materials, tools) are stored in warehouses and are used outdoors on construction sites. Warehouses are equipped with RFID readers, construction site workers are equipped with smartphones, and assets are tagged with RFID and BLE tags. If an asset leaves the warehouse, its RFID tag is scanned, and if a construction site worker passes by the asset in the field at close range, the BLE tag's RSS values are scanned by a mobile application installed on the worker's smartphone. With the knowledge of the RFID scanner's location and the smartphone's GPS coordinates, the approximate location of a scanned asset can be determined.

\subsubsection{UWB-based Approaches}

UWB is still not broadly available on mobile devices, with only a few flagship phones supporting it at the time of writing. However, few promising approaches are listed, such as \cite{leng2022design,grosswindhager2019snaploc,djosic2021fingerprinting}.

In high-precision, Line-of-Sight (LoS) tracking, UWB shows great potential thanks to its centimeter-level accuracy. \cite{djosic2021fingerprinting} investigates how UWB performs in complex indoor environments with partial or non-LoS connections available. Based on ToF ranging measurements, trilateration is used to position the node. In contrast, a fingerprinting-based algorithm provides additional context in cases of insufficient LoS measurements. The key aspect of their work is that fingerprints are not labeled with locations but rather with distances to a set of pre-defined reference points, which are then used in trilateration. Their findings include that UWB represents an effective alternative in-room identification with an accuracy of more than 95\%. Further, the system achieved comparable accuracy in cases of 2 instead of 3 LoS connections, making UWB deployments more attractive as the number of anchor points might be reduced.

\cite{leng2022design} proposes a framework for using and managing multiple IoT devices in a hospital. The framework is organized in layers. The sensing layer covers technologies such as UWB, BLE, WiFi to perform indoor localization functions, flow analysis, and fall detection, among other functions specific to hospitals (ECG). Data collected from devices is sent for processing to a cloud backend via WiFi that implements the functionality for these different services. One of the drawbacks of the proposed solution is that the use of several technologies in a single device/board, as proposed by the authors, becomes unfeasible for use at scale from an economic viewpoint and the mobility of PMDs.

\cite{grosswindhager2019snaploc} presents SnapLoc, a UWB-based indoor localization system that allows tracking an unlimited number of tags, in contrast to existing solutions that are limited in the number of supported tags. The authors rely on the Decawave DW1000 UWB chip and multiple anchors to detect UWB signals of tags, which can achieve a decimeter-level positioning accuracy with a 90\% error of 33.4 cm. While the accuracy of UWB is far superior to technologies such as Bluetooth, WiFi, and RFID, the technology is still relatively expensive and of limited availability. Using multiple anchors, as in Snaploc, makes the approach extremely accurate at a high acquisition and operating cost. 

\subsubsection{Other Approaches and Indoor Tracking Surveys}

As with a slightly different approach,  \cite{guo2021demo}\cite{rossi2013roomsense} propose EchoLoc and RoomSense, approaches using \textit{acoustic} responses to a chirp emitted by a smartphone. The system is based on the location-specific features captured in a chirp's acoustic response or echo and utilizes fingerprinting to learn these response patterns. While achieving high accuracy and low localization errors, these approaches are not feasible in the context of PMD-Track due to the rather small coverage area of the fingerprint (less than 1~m), which would translate into a high site survey effort.

\cite{dardari2015indoor} presents a \textit{survey} on indoor tracking covering a wide range of theories, methods, and technologies. This work has fundamental importance once it provides a mathematical formulation considering the types of methods used for each type of technology, such as which types of path-loss models can be used in geometric-related measurements and estimation of position based on Time-of-Arrival (ToA). In this sense, this work is important to provide a theoretical basis for tag localization based on reference beacons and smartphones.

\cite{mainetti2014survey} \textit{surveys} the state-of-the-art on enabling technologies on the different localization methods that different technologies can use. The paper highlights the characteristics, advantages, and disadvantages of different technologies ranging from RFID, through WiFi and UWB, to Bluetooth. Although it provides an overview from a technological viewpoint, some protocols listed are outdated, and Bluetooth is this paper's main one (listed in version v3.0).

\subsubsection{Discussion}

\begin{table}[]
\centering
\caption{Comparison of related work: technology, approach, and pros/cons}
\label{tab:related-work}
\resizebox{\columnwidth}{!}{%
\begin{tabular}{|l|l|l|l|l|}
\hline
\textbf{Work}                    & \textbf{Technology}                                       & \textbf{Approach} & \textbf{Pros}                                                                                                                          & \textbf{Cons}                                                                                                                                              \\ \hline
\cite{rodrigues2022pmd}          & BLE                                                       & n.a.              & \begin{tabular}[c]{@{}l@{}}- Proposal of PMD-Track\\ - Gateway-less solution\end{tabular}                                              & - No evaluation                                                                                                                                            \\ \hline
\rowcolor[HTML]{EFEFEF} 
\cite{rodrigues2021bluepil}      & BLE                                                       & Multilat.         & \begin{tabular}[c]{@{}l@{}}- General-purpose\\ - Use of RSSI to estimate\\ positioning\end{tabular}                                    & \begin{tabular}[c]{@{}l@{}}- Completely passive approach\\ - Poor accuracy\end{tabular}                                                                    \\ \hline
\cite{yoo2018real}               & BLE, WiFi                                                 & Multilat.         & \begin{tabular}[c]{@{}l@{}}- Designed specifically\\ for hospitals\\ - BLE beacons\end{tabular}                                        & \begin{tabular}[c]{@{}l@{}}- Use gateways\\ - Heavy use of  APs\\ for multilateration\end{tabular}                                                         \\ \hline
\rowcolor[HTML]{EFEFEF} 
\cite{canton2017bluetooth}       & BLE                                                       & Multilat.         & \begin{tabular}[c]{@{}l@{}}- Advanced location \\ algorithm using multilateration\\ and filtering\\ - Active BLE tracking\end{tabular} & \begin{tabular}[c]{@{}l@{}}- Use gateways\\ - Low accuracy\end{tabular}                                                                                    \\ \hline
\cite{adjei2020developing}       & BLE                                                       & Fingerp.          & \begin{tabular}[c]{@{}l@{}}- Real-time BLE tracking\\ - Use of smartphones to\\ synchronize locations\end{tabular}                     & \begin{tabular}[c]{@{}l@{}}- Limited mobility\\ - Expensive solution\end{tabular}                                                                          \\ \hline
\rowcolor[HTML]{EFEFEF} 
\cite{youn2007wlan}              & BLE, WiFi                                                 & Fingerp.          & \begin{tabular}[c]{@{}l@{}}- Real-time tracking\\ - Evaluated within a hospital\\ environment\end{tabular}                             & \begin{tabular}[c]{@{}l@{}}- Low accuracy\\ - Heavy use of APs\end{tabular}                                                                                \\ \hline
\cite{hakim2006passive}          & RFID                                                      & Multilat.         & \begin{tabular}[c]{@{}l@{}}- Real-time tracking tested\\ in a hospital's premises\\ - Widely adopted solution\end{tabular}             & \begin{tabular}[c]{@{}l@{}}- Low accuracy and reliability\\ due to the passive tags\\ - Use multiple RFID readers\end{tabular}                             \\ \hline
\rowcolor[HTML]{EFEFEF} 
\cite{bisio2016new}              & RFID, BLE                                                 & Multilat.         & \begin{tabular}[c]{@{}l@{}}- Advanced location algorithm \\ combining RFID and BLE\\ - High accuraccy\end{tabular}                     & \begin{tabular}[c]{@{}l@{}}- Use multiple RFID readers\\ - High battery consumption of\\ smartphones\end{tabular}                                          \\ \hline
\cite{djosic2021fingerprinting}  & UWB                                                       & Multilat.         & \begin{tabular}[c]{@{}l@{}}- High-accuracy\\ - Real-time tracking\end{tabular}                                                         & - Requires Line-of-Sight                                                                                                                                   \\ \hline
\rowcolor[HTML]{EFEFEF} 
\cite{leng2022design}            & \begin{tabular}[c]{@{}l@{}}UWB, BLE, \\ WiFi\end{tabular} & n.a.              & \begin{tabular}[c]{@{}l@{}}- Holistic framework integrating\\ IoT devices\\ - Hospitals' specific\end{tabular}                         & \begin{tabular}[c]{@{}l@{}}- Unfeasible due to the number \\ of sensors on a single board\\ - High use of AP's and external \\ cloud solution\end{tabular} \\ \hline
\cite{grosswindhager2019snaploc} & UWB                                                       & Multilat.         & - High-accuracy approach                                                                                                               & - Expensive hardware                                                                                                                                       \\ \hline
\rowcolor[HTML]{EFEFEF} 
\cite{guo2021demo}               & Acoustic                                                  & n.a.              & \begin{tabular}[c]{@{}l@{}}- High accuraccy\\ - Minimal additional infrastructure\\ requried\end{tabular}                              & \begin{tabular}[c]{@{}l@{}}- Low accuracy\\ - Low coverage\end{tabular}                                                                                    \\ \hline
\cite{rossi2013roomsense}        & Acoustic                                                  & n.a.              & \begin{tabular}[c]{@{}l@{}}- High accuracy\\ - Low error\end{tabular}                                                                  & \begin{tabular}[c]{@{}l@{}}- Low accuracy\\ - Low coverage\end{tabular}                                                                                    \\ \hline
\rowcolor[HTML]{EFEFEF} 
\cite{dardari2015indoor}         & All                                                    & All               & \begin{tabular}[c]{@{}l@{}}Indicate a wide range of theories, \\ methods, and technologies\end{tabular}                                & \begin{tabular}[c]{@{}l@{}}Lack of analysis on practical \\ (real world) approaches\end{tabular}                                                           \\ \hline
\cite{mainetti2014survey}        & All                                                    & All               & \begin{tabular}[c]{@{}l@{}}Present enabling technologies \\ for indoor tracking in real-world \\ scenarios\end{tabular}                & \begin{tabular}[c]{@{}l@{}}Overview based on BL v3.0.\\ As of now, BLE is in v5.2\end{tabular}                                                             \\ \hline
\end{tabular}%
}
\begin{itemize}
\item[] \textit{n.a.: not available; multilat.: multilateration; fingerp.: fingerprinting}
\end{itemize}
\end{table}

Table \ref{tab:related-work} provides an overview of related work, comparing their major characteristics, such as used technology and approaches, and the pros and cons of their applicability to the proposed solution's scope. UWB-based solutions' precision is higher than BLE, WiFi, and RFID solutions. However, UWB solutions are not as mature and pervasive as others, which impacts hardware availability and costs. Once hospitals are equipped with thousands of portable devices, the relation between cost and benefit is often an important requirement in adopting tracking solutions.

In this regard, the advantage of acoustic-based approaches is that they require minimal additional infrastructure. For example, additional gateways/readers and APs are not required to operate with such solutions. However, they are not widely utilized in practice, and their limited coverage area renders them impractical for large-scale deployments. In addition, it is unclear how environmental changes and noise affect their localization performance.

Table \ref{tab:related-work} also shows that solutions based on multilateration of signals are in the majority, which does not necessarily imply greater precision. In this case, solutions based on multilateration are simpler to calculate since only three overlapping reference points are needed to determine the position of a device. In contrast, fingerprinting requires a longer training period with signal collection at several points to statistically estimate the probability that a detected signal is close to one or more reference points.

In indoor environments, where LoS connections are not guaranteed and triangulation-based approaches fail, fingerprinting has become a popular choice among RSSI-based methods. Despite these benefits, a labor-intensive site survey is always required before large-scale deployment.
There have been numerous efforts to streamline the site survey procedure, some of which are entirely passive. Lastly, other approaches have focused on algorithms to improve the recognition and accuracy of fingerprints.

\section{PMD-Track Design}
\label{sec:design}

Traditional asset-tracking solutions track tagged assets by installing static gateways throughout the building. The contribution of PMD-Track is the absence of such gateways by leveraging employee workforce mobility. In PMD-Track, hospital staff, typically equipped with smartphones, move throughout the building daily, covering a large area of the hospital's premises and the gateway application is a BLE tracking software installed on the employee's smartphones that detects BLE beacons in proximity. Design considerations and requirements on an abstract level are:
\begin{itemize}
    \item \textbf{Accuracy}: the goal of the proposed solution is tracking PMDs within hospitals or healthcare facilities, which are segmented into rooms. Thus, room-level tracking is required.
    \item \textbf{Heterogeneous smartphones}: Deployed as a crowdsourced application, smartphone brands and models may be heterogeneous and certain limitations may be experienced in the reception of BLE signals or the usage of system resources. Hence, device heterogeneity must be taken into consideration.
    \item \textbf{Real-time location data}: Near real-time or fresh data is crucial for applications such as PMD-Track. Outdated information can lead to inefficient processes and wasted resources, leading to mistrust of users against the system. Thus, it is crucial to manage the user's expectations by communicating the staleness of information.
    \item \textbf{Privacy}: From a user perspective, the person's privacy carrying the smartphone mustn't be violated. Thus, it is crucial that the user's location is not tracked or cannot be inferred from other data. Furthermore, hospitals often implement strict IT-security policies. Thus, minimal friction and touch points with existing IT operations, such as connectivity services through an existing WiFi infrastructure, are needed.
\end{itemize}

\begin{figure}[!t]
    \centering 
    \includegraphics[width=1\linewidth,keepaspectratio]{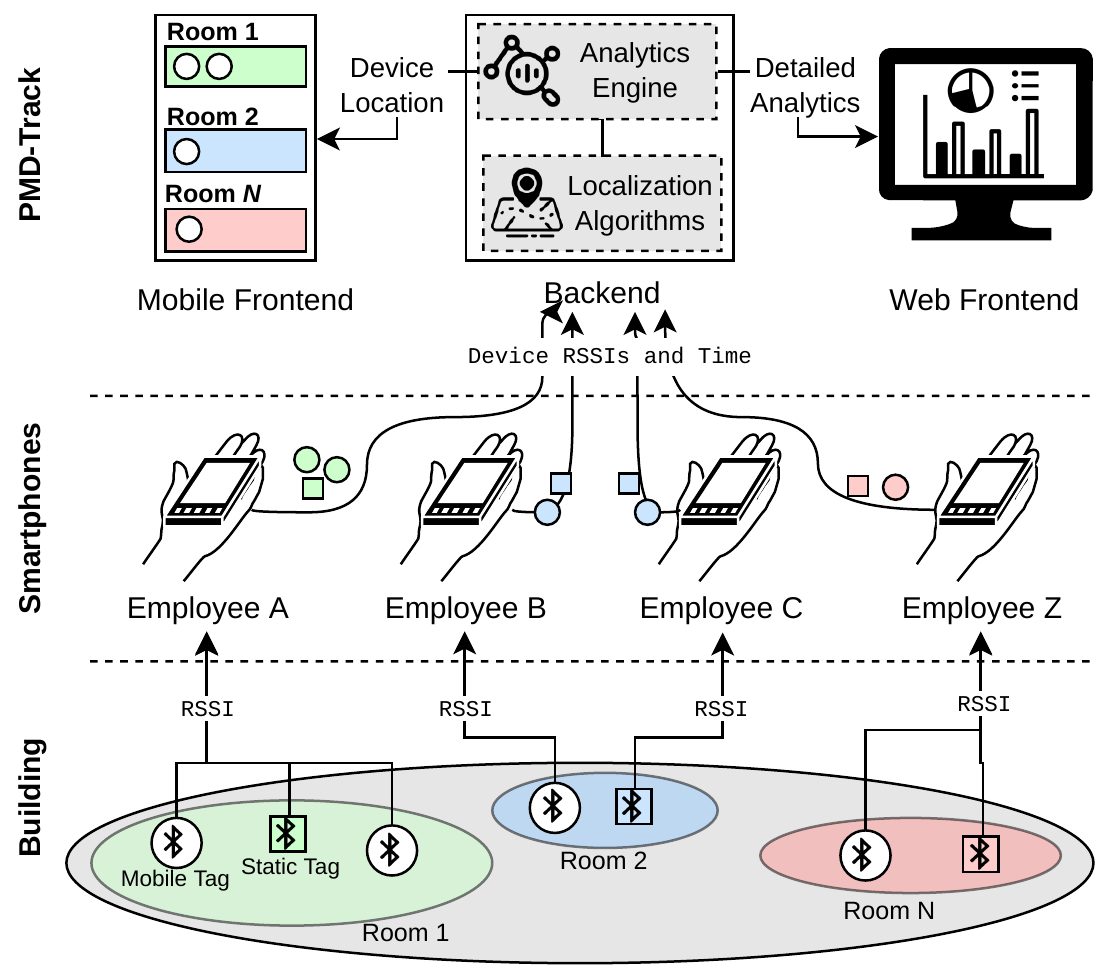}
    \caption{PMD-Track beacon scanning \cite{rodrigues2022pmd}}
    \label{fig:beacon-scanning}
\end{figure}

As illustrated in Figure \ref{fig:beacon-scanning}, smartphones interact - as the medical staff performs their daily activities - via BLE with BLE tags attached PMDs, and static BLE tags used as reference points. As soon as smartphones are in range with a BLE tag (static or mobile), it transmits the Received Signal Strength Indicator (RSSI) to a backend service via WiFi, which calculates the position of the PMD and updates its location in the inventory.  

The design consists of (1) a mobile application that records Bluetooth encounters between a mobile device and BLE tags placed on objects (represented by geometric shapes) and (2) a backend that receives information sent by smartphones in the form of message streams and uses it to calculate the approximate location (\eg Room 1, Room 2, Room N) of the object in real-time based on three primary reference points: the mobile device, a static tag, and a mobile tag.

After gathering information about static and mobile tags, the smartphone application timestamps their receipt and sends it to the backend service. At this stage, it is critical to minimize excessive processing power and data transfer by sending data in rolling time periods, hence minimizing smartphone battery depletion.

Upon receiving data, the backend service \textit{(1)} executes the localization algorithm and updates the PMD's location by triangulating BLE RSSI values, using the known position of static (\ie anchor) BLE tags as a reference, and \textit{(2)} calculates metrics regarding their usage (\eg heatmap of PMD by type, their time spent in rooms, or the number of times a PMD has been moved). The analytics engine provides data for two frontends: a mobile (simple, intuitive localization of PMDs) and a Web-based (PMD-specific detailed analytics) one.

\begin{figure}[!h]
    \centering 
    \includegraphics[width=1\linewidth,keepaspectratio]{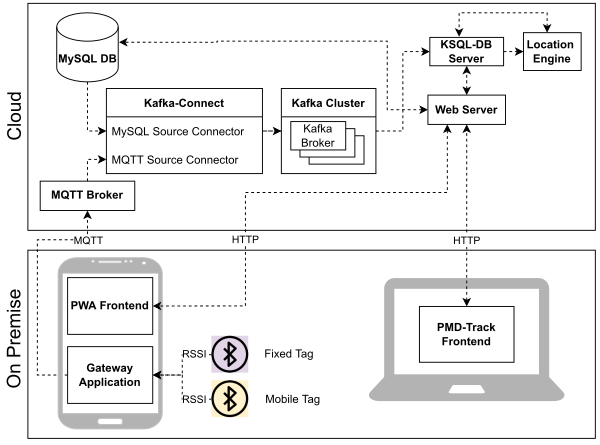}
    \caption{PMD-Track Architecture \cite{pmdExperience2023}. PWA: Progressive Web Application}
  \label{fig:architecture}
\end{figure}

Figure \ref{fig:architecture} illustrates PMD-Track's architecture overview, which comprises the following components summarized below and further described in the following subsections:
\begin{itemize}
    \item \textbf{MySQL DB}: Store static inventory data
    \item \textbf{Kafka-connect and cluster}: Integrate various data sources with the Kafka Cluster instance \cite{apacheKafka}
    \item \textbf{KSQL-DB}: Database to query and aggregate streams
    \item \textbf{Location engine}: Predict location of PMDs
    \item \textbf{Web server}: Exposes location data through REST API
    \item \textbf{MQTT broker}: Facilitates on-premise to cloud communication
    \item \textbf{PWA frontend}: Display asset data on mobile \cite{fortunato2018progressive}
    \item \textbf{PMD-Track frontend}: Display asset data on desktop
    \item \textbf{Gateway application}: Beacon scanning application
    \item \textbf{Fixed/mobile tag}: BLE beacons (fixed location / on asset)
\end{itemize}

\subsection{BLE Beacons}
BLE beacons are small, inexpensive, and battery-powered devices emitting a periodic BLE signal to advertise their presence to nearby devices. A beacon is uniquely identifiable through its Medium Access Control (MAC) address and RSSI value, and the distance between it and the receiving device can be approximated using the path-loss model. A common application scenario is proximity detection, where an application or device is scanning for beacons in its perimeter to establish a geospatial context associated with the scanned beacons. In the case of PMD-Track, two types of BLE beacons are distinguished. First, \textbf{mobile} beacons are attached to the assets one wants to track and are moving along with them. Their MAC address is associated with a specific piece of inventory. Second, \textbf{fixed} beacons are installed in predefined areas of the building and thus do not change their location. Their purpose is to serve as reference points for determining the current location of the scanning device within the building. Locating the scanning device allows the inferring of the location of assets it has detected in its vicinity.

\subsection{Gateway Application}
Traditional asset-tracking solutions rely on installing static gateways throughout the building to track tagged assets. A key requirement and novelty of PMD-Track is the absence of such a static gateway by leveraging the mobility of the employee workforce. Typically equipped with a smartphone, hospital staff moves through the building, covering a wide area of the hospital's premises daily. The gateway application is a BLE tracking software installed on the employee's smartphones that detects BLE beacons in proximity. The appplication runs passively and in the smartphone's background without requiring user interaction; thus, employees are not disturbed in their activities. It scans mobile and fixed BLE beacons in range and relays the timestamped information to a cloud service as illustrated in Figure \ref{fig:beacon-scanning}. Reliable and uninterrupted operation of the gateway application is therefore of utmost importance to provide up-to-date information.

\begin{figure}[t]
    \centering 
    \includegraphics[width=1\linewidth]{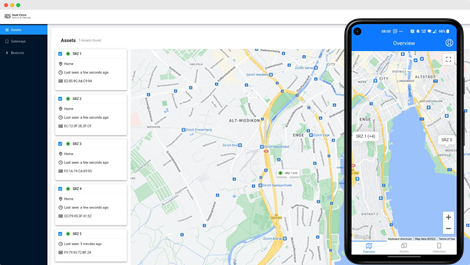}
    \caption{Web and PWA mobile frontends}
  \label{fig:frontend}
\end{figure}

\subsection{Frontend views}
Information about inventory location is displayed through two User Interfaces (UI) as depicted in Figure~\ref{fig:frontend}. Field workers are provided with a mobile application to quickly query and locate tracked assets. It is important to note that the mobile UI application is separate and not integrated into the existing gateway application that also runs on the smartphone. The reason for this separation is two-fold. First, apps that display information to users can efficiently be built by leveraging cross-platform frameworks to reduce development effort.
Conversely, the gateway application requires a native implementation, as accessing the smartphone's hardware resources (\ie BLE scanning in the background) requires access to OS-level libraries to acquire necessary permissions and to ensure continuous background operation. Second, due to privacy concerns, a separate UI application allows users to consume asset location information without opting in on the BLE tracking. Apart from the mobile application, a web-based dashboard provides asset location information in a single view, suitable for larger displays, such as desktops.

\subsection{Communication between gateway and cloud}
 Message Queuing Telemetry Transport (MQTT) is a lightweight, publish-subscribe, machine-to-machine communication protocol \cite{mqtt}. Its low resource consumption makes it a popular choice for IoT applications in resource-constrained environments. In PMD-Track, MQTT facilitates communication between the gateway application and the cloud service. The gateway application connects to the MQTT broker and publishes aggregated BLE scan results on a predefined topic in regular intervals (\eg every minute). Downstream, a consumer application can connect to the broker, subscribe to the topic, and process the received messages.

The messages on the MQTT topic form a continuous stream of events, each representing a device's encounters between the gateway (\ie smartphone) and a BLE beacon it has recently scanned - termed \textit{BleScanEvent}. Its properties include a \textit{client\_id} identifying the gateway application, a \textit{mac\_address} associated with the BLE beacon, its \textit{rssi} value, and a \textit{timestamp} when it was detected. These messages are fed into a stream processing framework by a connector application that subscribes to the MQTT topic and connects to the stream processing framework to forward the messages on a dedicated stream. Client applications can then consume, aggregate, and act on those events to either derive state or emit new events.

\subsection{Streaming and static data processing}

Apache Kafka has been chosen as a stream processing framework to facilitate the communication of services and to enable the processing of BLE scan results \cite{apacheKafka}. For simplicity, the Kafka infrastructure scaling has been kept at a minimum with single cluster and single partition deployment. To integrate the MySQL database and the MQTT broker with the Kafka stream processing framework, Kafka-Connect, a free and open-source component of Confluent's Kafka suite, has been chosen.

The \textit{BleScanEvents} received in the cloud are small bits of information from various data sources, depending on how many gateway applications are running. There is no guarantee that, eventually, all BLE beacons will be scanned after a certain time, as the gateway only detects what currently is in range. If a beacon is lost or never comes into range of a gateway, this beacon will never be detected; thus, no \textit{BleScanEvent} will ever be emitted in the system. In the context of asset tracking, the detection of the absence of an asset is equally important as detecting its presence. Hence, the need for storing a predefined, stable list of inventory as a base data set occurs. 

Storing a list of inventory can be achieved with a variety of proven technologies. In PMD-Track, a relational database stores a table of BLE beacons along with their type (\ie \textit{mobile}, \textit{fixed}) and their \textit{mac\_address}. Similarly, a table stores a list of active gateway applications along with their \textit{client\_id} and human-readable \textit{name}. Finally, a table stores room information about the building. To make the static information available in the stream processing framework, a connector application periodically fetches the information from the relational database and feeds it into a dedicated stream. Serving as the base data set, a new, enriched stream joins \textit{BleScanEvents} with the static beacon data from the relational database, filtering out non-inventoried beacons.

\subsection{Location Engine}

Room-level positioning can be achieved using different localization algorithms, each with advantages and disadvantages. The location engine receives a stream of enriched \textit{BleScanEvents} and produces a stream of positioned \textit{mobile} beacons associated with a location (\ie room). Thus, the asset associated with the \textit{mobile} beacon has been located in the given room. 

\subsubsection{Data Preparation}

\begin{figure}[t]
  \centering
  \includegraphics[width=1\linewidth,keepaspectratio]{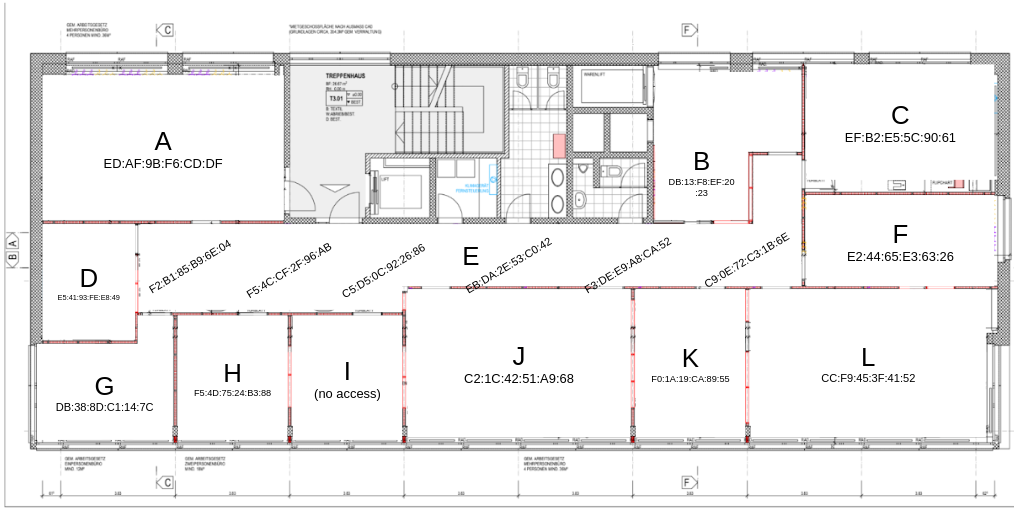}
  \caption{Floor plan office at Europastrasse Zürich}
  \label{fig:floorplan_europastrasse}
\end{figure}

Preparing a single data set per test location (\cf Figure \ref{fig:floorplan_europastrasse}) is a fundamental step for the subsequent model training and comparison of the two algorithms. The following steps for data gathering and preparation have been done in both test locations. Each room, labeled with a capital letter (A-Z), is equipped with a \textit{fixed} BLE beacon, denoted by the MAC address below the letter of the room. The beacons are installed in the center of the ceiling of each room, and the mapping between the room label and MAC address is stored.

\begin{figure}[h]
  \centering
  \includegraphics[width=.5\linewidth,keepaspectratio]{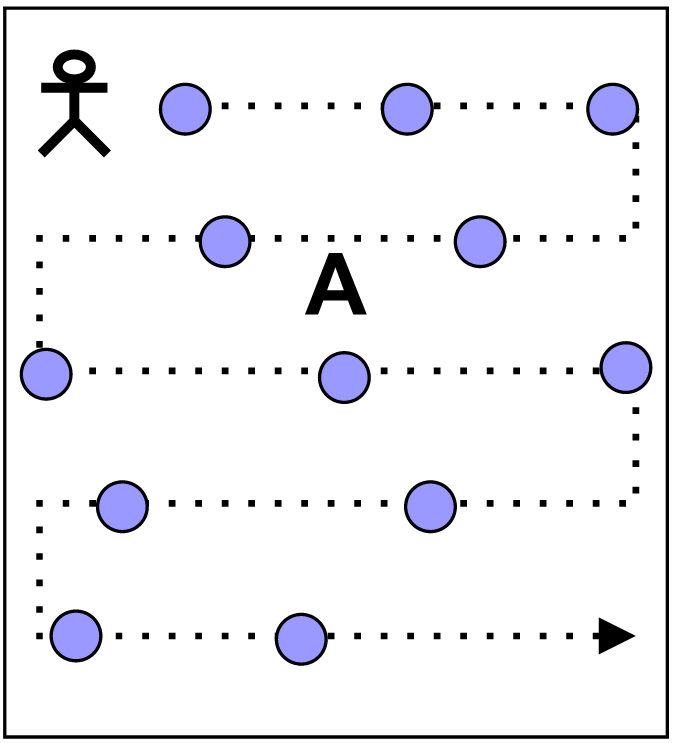}
  \caption{Collecting fingerprints at different positions within a room}
  \label{fig:collecting_fingerprints}
\end{figure}

After preparing the test location with beacons, the data set can be gathered. In this process, a person visits each room and takes a certain amount of RSSI samples of beacons that are in range within that room. Recording these RSSI samples is done using a Smartphone app. The user enters the room label where the sampling occurs, and the app performs repeated BLE scans. Within each scan --- which only lasts a few seconds --- the app records the MAC address and RSSI value of the \textit{fixed} beacons it detected. At the end of each scan, the detected beacons and RSSI values are stored in a new data point along with the room label. To make sure RSSI samples are not only taken at a single position within the room, the person moves throughout the room while collecting the samples, as seen in Figure \ref{fig:collecting_fingerprints} 

Once 1,000 data points are collected within the same room, the person stops the app and moves to the next room, where the procedure is repeated. Repeating this for \(k\) rooms yields the final data set of \(k * 1,000\) rows and \(k+1\) columns, as one column accounts for the room label. As the data collection is terminated after a fixed number of measurements, the final data set is balanced by containing an equal number of samples per class/room. Table \ref{tab:dataset} shows an example of the final data set.

\begin{table}[t]
\centering
\caption{Data set}\label{tab:dataset}
\resizebox{\columnwidth}{!}{%
\begin{tabular}{|l|l|l|l|l|}
\hline
\textbf{BEACON\_A} & \textbf{BEACON\_B} & \textbf{...} & \textbf{BEACON\_L} & \textbf{Room} \\ \hline
-65       & -70       &     & -99       & A    \\ \hline
\vdots & \vdots & & \vdots & \vdots\\ \hline
-99       & -75       &     & -65       & L \\ \hline
\end{tabular}
}
\end{table}

The RSSI value captures the distance relationship between the beacon and the smartphone. Due to environmental noise, not every beacon might be captured in a data point. This results in the sparsity of the final data set (\ie cells being null). As fingerprinting and multilateration-based approaches cannot operate on sparse input data, two imputation strategies are applied to fill in missing observations. A certain beacon might never be detected for a given class in the first case. Considering the floor plan in Figure \ref{fig:floorplan_europastrasse}, the data points for room \(A\) might not have any records of the beacon in room \(L\) since it is too far away and, thus, out of range. In this case, the column of beacon \(L\) for room \(A\) is set to the constant value of -200 to indicate an out-of-range beacon. 

\begin{table}[htbp]
\centering
\caption{Value imputation}\
\begin{tabular}{|l|l|l|}
\multicolumn{3}{c}{Fixed value imputation} \\ \hline
\textbf{BEACON\_J} & \textbf{BEACON\_J (imputed)} & \textbf{Room} \\ \hline
null & -200      & A    \\ \hline
null & -200      & A    \\ \hline
null & -200      & A    \\ \hline
null & -200      & A    \\ \hline
\multicolumn{3}{c}{Column mean imputation} \\ \hline
-65  & -65     & A    \\ \hline
null & -64.67  & A    \\ \hline
-62  & -62     & A    \\ \hline
-67  & -67     & A    \\ \hline
\end{tabular}
\end{table}

RSSI values are only partially absent in the second case due to shadowing or other signal interference effects. Considering the floor plan in Figure \ref{fig:floorplan_europastrasse}, rows might exist for room \(A\) where the RSSI value of the beacon of the adjacent room \(D\) is missing. In this case, the conditioned mean of beacon \(D\) given room \(A\) is used to impute the missing observations. \(value = mean(D | Room=A)\). Applying the described imputation strategy for each room eventually yields a non-sparse data set.

\subsubsection{Model Training}

Once missing values have been imputed and the data set is complete, the two models can be defined, fitted, and evaluated. The classification task of predicting the room label given a set of RSSI measurements is typical. A k-nearest neighbors (kNN) classifier is utilized for the fingerprinting model. In contrast, a multilateration algorithm is modified to approximate the room using geometric calculations (as described in Section \ref{sec:fundamentals}).

\begin{figure}[h]
  \centering
  \includegraphics[width=1\linewidth]{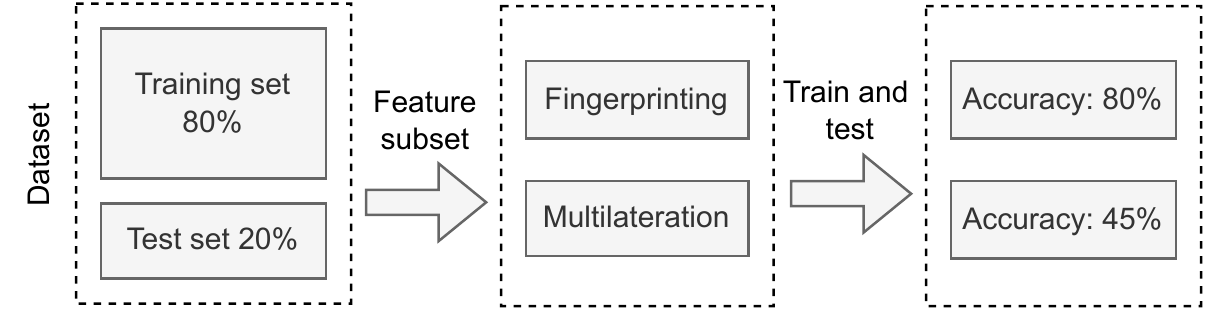}
  \caption{Steps taken in the model training}
  \label{fig:model-training}
\end{figure}

Figure \ref{fig:model-training} presents the stages for training the model, including preparing the dataset described in the previous section. The dataset is first partitioned into a training and testing set with an 80\% and 20\% split, respectively. Subsequently, and in the case of kNN, the model fits on the training set. 
\begin{itemize}
    \item \textbf{Multilateration} localization yields a geospatial position with $x$ and $y$ coordinates. In non-LoS conditions, determining the distance based on the RSSI value is error-prone due to environmental signal interference. Further, in the case of PMD-Track, room-level granularity suffices compared to an exact position. Thus, the design of an adapted multilateration algorithm predicting the room label instead of coordinates in combination with floor plan information is key.
    \item \textbf{kNN} uses distance-based metrics to determine neighboring data points. In case of the collected RSSI values (as described in Section \ref{sec:fundamentals}), the data set is split into a disjoint training and test data set with the dimensions \(cn \times{k+1}\) and \((1-c)n \times{k+1}\) for a train-test-split coefficient \(c\) (\ie c=0.2). In the case of PMD-Track, the train-test-split was set to 0.2, \(k\) was set to 7, and \(l_2\) norm (Euclidean) was used as the distance metric.    
\end{itemize}

\subsubsection{Adapted Multilateration}
Distance measurements might not be accurate due to noise in the RSSI signal. Thus, an overlapping intersection area might exist rather than a deterministic intersection point between the ranges of three known points. Overlaying this intersection area with floor plan information yields coverage areas on a room-level basis. The room with the largest coverage area is considered the most probable current location of the asset. The following sections describe different localization scenarios on a case-by-case basis and how they are resolved to a room-label prediction.

\begin{figure}[h]
  \centering
  \includegraphics[width=1\linewidth,keepaspectratio]{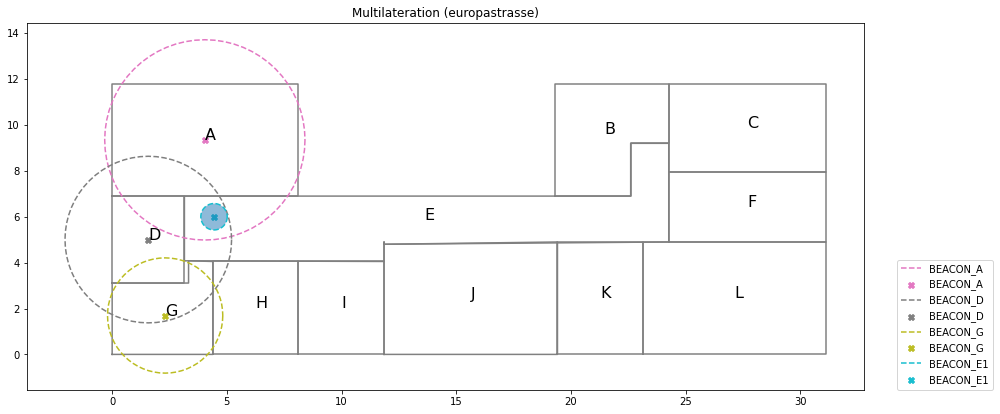}
  \caption{Multilateration: close proximity}
  \label{fig:multi_case_1}
\end{figure}

The \textbf{first case} handles the situation when the smartphone and the fixed beacon are nearby so that they can be considered in the same room. This is the case for strong RSSI values above -70 dBm, which translates to a range of up to approximately 2 meters. Figure \ref{fig:multi_case_1} illustrates a situation where the smartphone detects fixed beacons from rooms A, D, G, and E. The radius of the dotted circles is calculated based on the path-loss model and indicates the distance measured from the smartphone to each of them. As seen from the plot, the beacon in room E (hallway) is nearby and, in this case, within the range of 2 meters. Thus, the predicted room label in this situation is E. If no beacon is within the range of a 2-meter radius, the intersection area of the detected fixed beacons is~calculated.

\begin{figure}[h]
    \centering
    \includegraphics[width=1\linewidth,keepaspectratio]{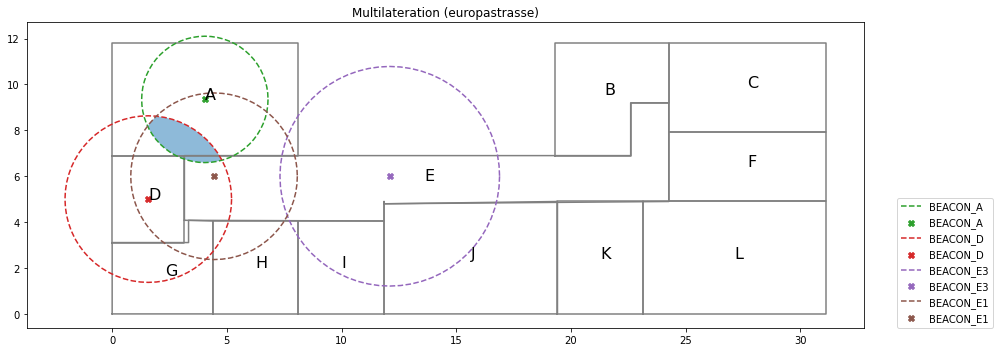}
    \caption{Multilateration: Multiple intersections, max intersection cardinality}
    \label{fig:multi_case_3-Design}
\end{figure}

Multiple beacons are detected in the \textbf{second case}, and multiple intersection areas might exist. Figure \ref{fig:multi_case_3-Design} shows such a situation where beacons A, D, E1, and E3 are detected with their respective intersection areas. To characterize different intersection areas, the concepts of \textit{intersection set} and \textit{intersection cardinality} are introduced. Given an intersection \(i\), a set of origin shapes can be defined as the intersection set \(s_{inter}(i)\) whose elements yield the intersection \(i\). For example, given two circles \(c_1\) and \(c_2\) that produce intersection \(i\), the intersection set is defined as \(s_{inter}(i) = \{c_1, c_2\}\). Given an intersection set \(s\), the intersection cardinality is defined as \(\parallel s \parallel\). In case of the previous example, \(\parallel s_{inter}(i) \parallel  =  \parallel \{c_1, c_2\} \parallel = 2\).

Returning to the situation shown in Figure \ref{fig:multi_case_3-Design}, one can observe that the intersection of beacons A, D, and E1 has the highest cardinality. \ie 
\[\exists s_{inter}(i) \:| \parallel s_{inter}(i) \parallel > \parallel s_{inter}(j) \parallel \forall j \in intersec \wedge i \neq j\]

Analogously to the previous case, once the intersection with the highest cardinality is determined, the intersection area is overlaid on the floor plan, and the room with the largest coverage area is predicted as the room label. In this case, it is room A. As shown in the previous example, multiple intersections with different cardinalities may occur. If multiple intersections exist with the same maximum cardinality, another selection criteria is applied to determine the intersection to be overlaid on the floor plan. To this end, the notion of \textit{radii sum} is introduced. Given an intersection set \(s_{inter}(i) = \{c_1, c_2, ..., c_n\}\), the radii sum is defined as the sum of the individual radii \(r_{c_k}\)
\[R_i = \sum_{k=1}^{n} r_{c_k}\]

\begin{figure}[h]
    \centering
    \includegraphics[width=1\linewidth,keepaspectratio]{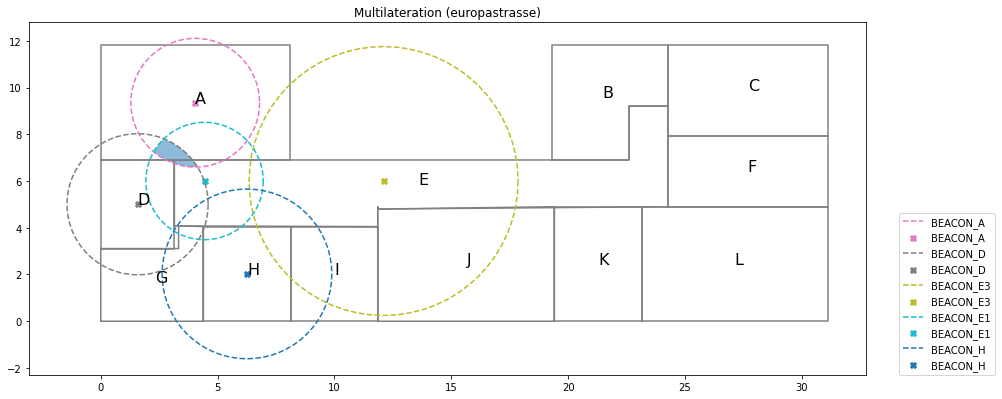}
    \caption{Multilateration: Multiple intersections, min sum of radii}
    \label{fig:multi_case_4}
\end{figure}

Such a situation of the \textbf{third case} is depicted in Figure \ref{fig:multi_case_4} where there exist three intersections with cardinality 3 (\(\{A, D, E1\}\), \(\{D, E1, H\}\), \(\{H, E1, E3\}\)). Because RSSI noise increases with distance, a strong RSSI signal is more stable and accurate than a weak signal. The radius of the dotted circles inversely correlates to the signal strength (\ie the stronger the RSSI signal, the smaller the radius). Thus, the smaller circles provide a more accurate indication of the distance than larger circles. On this premise, the intersection set \(\{A, D, E1\}\) is considered the most reliable intersection to be used for room label prediction. Overlaying it on the floor plan reveals that room A has the largest coverage area.

\begin{figure}[h]
    \centering
    \includegraphics[width=1\linewidth,keepaspectratio]{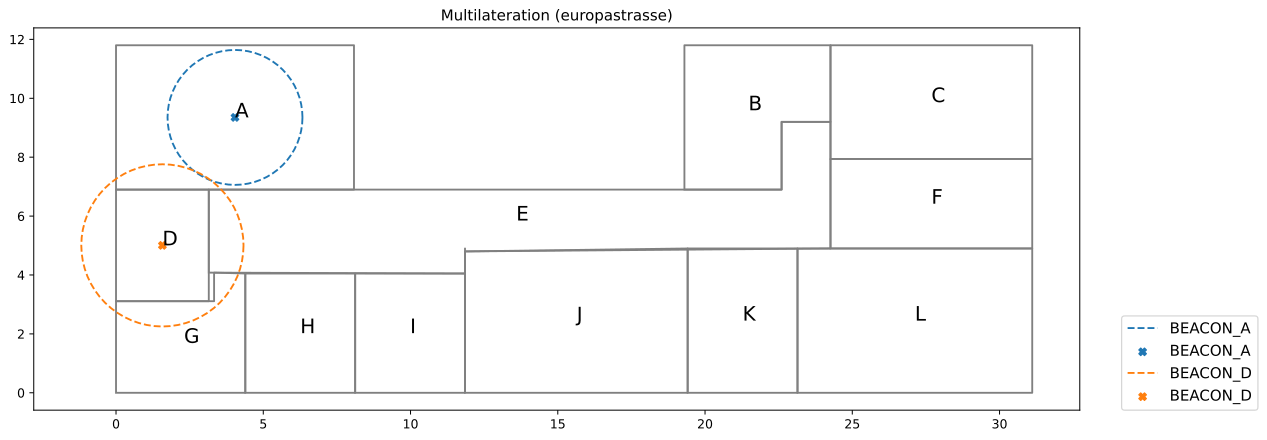}
    \caption{Multilateration: No intersection}
    \label{fig:multi_case_5}
\end{figure}

\begin{figure*}[!ht]
    \begin{subfigure}{}
        \includegraphics[width=.49\linewidth]{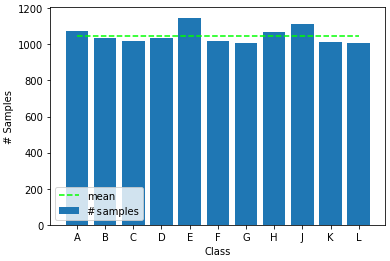} 
    \end{subfigure}
    \begin{subfigure}{}
        \includegraphics[width=.49\linewidth]{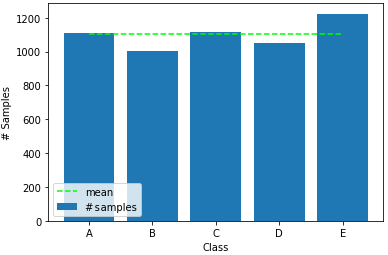}
    \end{subfigure} 
    \caption{Evaluation environments considering the number of samples per class: (Left) Office at Europastrasse and (Right) Apartment at Waffenplatz}
    \label{fig:exp_env}
\end{figure*}

Finally, the \textbf{fourth case} stems from a situation where no two beacon signals overlap. Given that there is no overlap between static beacons, the smartphone, and the mobile beacon associated with the PMD, this is the worst-case scenario. In this instance, the conventional multilateration method would fail because it does not account for historic data and does not perform approximations without a minimum overlap of three beacons. This case is illustrated in Figure~\ref{fig:multi_case_5} where the beacon A and D ranges do not overlap or touch at any point. In this case, the beacon closest to the smartphone is considered to predict the room label. In the situation, the room predicted is A as beacon A has a stronger RSSI signal and is thus closer to the smartphone.

\section{Evaluation}
\label{sec:evaluation}

Experiments were independently performed in two test locations: an apartment building for initial experiments and configurations and an office building with a similar floor plan to a hospital floor. Importantly, permission to conduct experiments in a hospital environment has not been granted. For both test locations, an initial data set was gathered on which fingerprinting, and multilateration-based approaches were trained and evaluated. The experiments were based on i10 Durable Beacon \cite{i10Beacon} as static beacons, E7 Plus beacon \cite{e7Beacon} as mobile beacons, and Android phones hosting the gateway application. The following experiments were conducted to analyze the tracking model accuracy:
\begin{itemize}
    \item Accuracy vs number of beacons (Subsection \ref{subsub:accuracy-vs-numBeacons})
    \item Accuracy vs placement of beacons (Subsection \ref{subsub:accuracy-vs-placement}) 
    \item Accuracy vs training size (Subsection \ref{subsub:accuracy-training}) 
    \item Economic analysis (Subsection \ref{subsubsec:economic})
\end{itemize}

The goal is to analyze and compare the model accuracy under different input feature configurations to minimize the feature input space (\ie number of beacons) while maximizing model accuracy. A column in the data set corresponds to a certain beacon or feature. As the initial data set was gathered with one beacon per room, training and evaluating the model with different beacon configurations can be achieved by simply considering a subset of the columns (\ie beacons or features) at a time. Lastly, this section provides considerations on the limitations and lessons learned (\cf Subsection \ref{subsub:lessons-and-limitations})

\subsubsection{Model Accuracy vs. Number of Beacons}
\label{subsub:accuracy-vs-numBeacons}

The cost of installing beacons can be a significant barrier to implementing a tracking solution such as PMD-Track. Thus, evaluating the trade-off between model accuracy and the number of beacons is essential to find an optimal number that is economically viable (economic analysis is presented in Subsection \ref{subsubsec:economic}), straightforward to be maintained, and that can be scaled to meet a variety of evolving needs. Also, specific aspects of each beacon, such as coverage area and calibration, require careful consideration as these influence the model's accuracy.

Input feature combinations are called beacon subsets in our model. Thus, the model's accuracy analysis under different input feature combinations is needed, and all data classes/rooms in the testing scenarios are evaluated. As a non-skewed data measure, accuracy was utilized. Figure~\ref{fig:exp_env} depicts the number of samples per class, demonstrating that both data sets are well-balanced concerning the class members. 

\begin{figure}[t]
    \centering
    \includegraphics[width=1\linewidth,keepaspectratio]{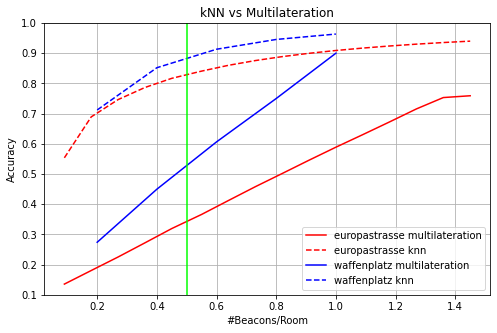}
    \caption{Accuracy threshold}
    \label{fig:accuracy_vs_beacons_per_room}
\end{figure}

\begin{figure*}[ht]
    \begin{subfigure}{}
        \includegraphics[width=.475\linewidth]{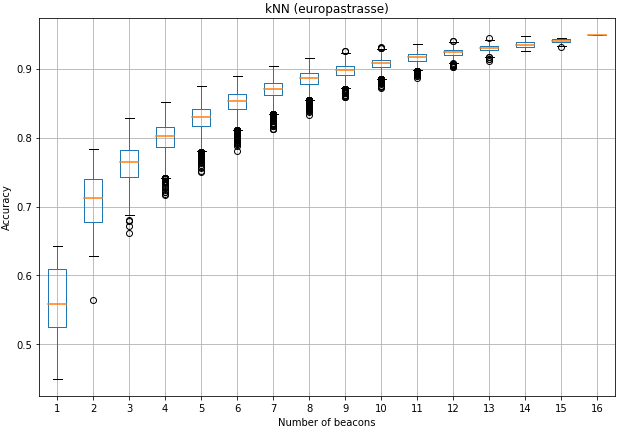} 
    \end{subfigure}
    \begin{subfigure}{}
        \includegraphics[width=.475\linewidth]{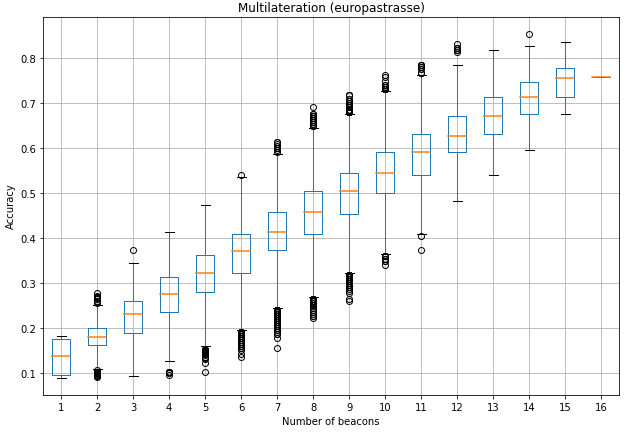}
    \end{subfigure} 
    \caption{Accuracy box plot grouped by number of features (Europastrasse): (Left) kNN fingerprinting and (Right) Multilateration}
    \label{fig:boxplot-europastrasse}
\end{figure*}

To evaluate the accuracy of a specific input feature combination, the data set is first partitioned into a training and testing set with an 80\% and 20\% split, respectively. Subsequently, and in the case of kNN, the model fits the training set. Finally, the accuracy is calculated on the testing set by computing the number of correctly classified samples divided by the total number of samples in the testing set. Cross-validation is performed to achieve a more reliable accuracy score by repeatedly training and testing the model on different train-test-splits.

As mentioned above, a feature combination is a subset of the total number of features. Thus, there are feature combinations of length \(1, 2, ..., d\) where \(d\) is the number of available features. Performing this evaluation is computationally expensive due to the high number of combinations. 
\[combinations = \sum_{i=1}^{d} \frac{d!}{i! (d - i)!}\]
The office and apartment test locations yield 65,535 and 31 combinations, respectively. The following analysis groups accuracy scores by the average number of features involved.

In addition, Figure \ref{fig:accuracy_vs_beacons_per_room} shows accuracy on the y-axis and the number of beacons per room ratio on the x-axis. For example, the green vertical line indicates the accuracy score for a beacon per room ratio of 0.5. \textit{For the office test location, equipping half of all available rooms with beacons yields an accuracy of 35\% and 83\% on average for the multilateration-based and kNN-based approaches, respectively}. It is worth noting that the office test locations are equipped with 16 beacons and have 11 rooms, which explains the beacon-per-room ratio of up to 1.4. Based on this interpretation, one can easily observe that kNN performs considerably better than multilateration for all beacon-per-room ratios and across test locations. Further, it appears that accuracy for the fingerprinting-based approach grows logarithmically, whereas the multilateration-based approach has a more linearly-shaped growth. 

\begin{figure}[t]
    \centering
    \includegraphics[width=1\columnwidth,trim={0 0 0 0.75cm},clip]{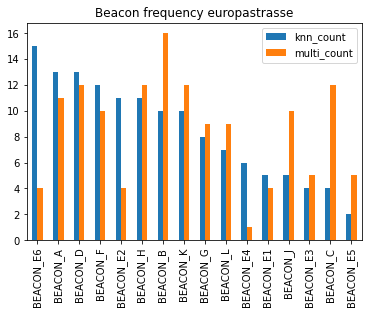}
    \caption{Beacon frequency - Office at Europastrasse}
    \label{fig:beacon_frequency_europa}
\end{figure}

The box plot in Figure \ref{fig:boxplot-europastrasse} shows the accuracy distribution within a group of features. A wide distribution indicates that the choice of beacons has a high impact on the accuracy, whereas a narrow distribution suggests that the choice is less relevant concerning achieved accuracy. For example, in Figure \ref{fig:boxplot-europastrasse} for the office test location and the kNN model, one can observe that installing 3 beacons in arbitrary rooms yields a median accuracy of approximately 77\% with Q1 and Q2 quartiles around 75\% and 78\% respectively. It can be observed that the accuracy distributions get narrower as the number of beacons increases. Alternatively, for a low number of beacons (1-3), the choice of beacons seems to have a high impact, as seen by the wider distributions. 

\begin{figure*}[!htbp]
    \begin{subfigure}{}
        \includegraphics[width=.48\linewidth,trim={0 0 0 0.75cm},clip]{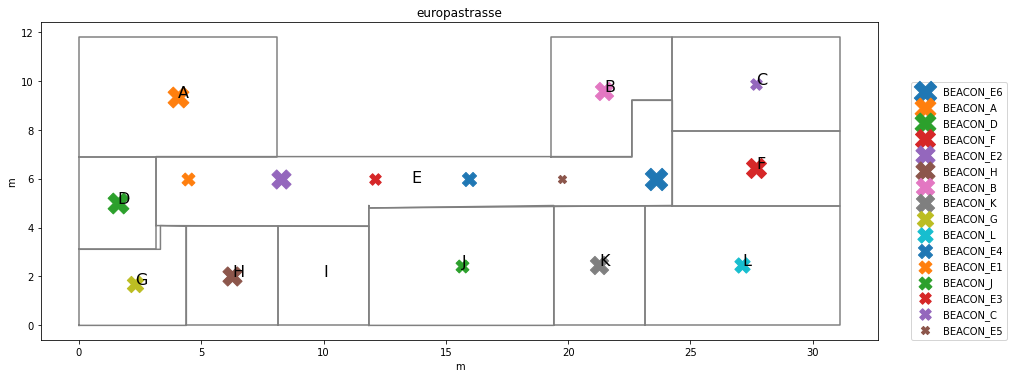} 
    \end{subfigure}
    \begin{subfigure}{}
        \includegraphics[width=.48\linewidth,trim={0 0 0 0.75cm},clip]{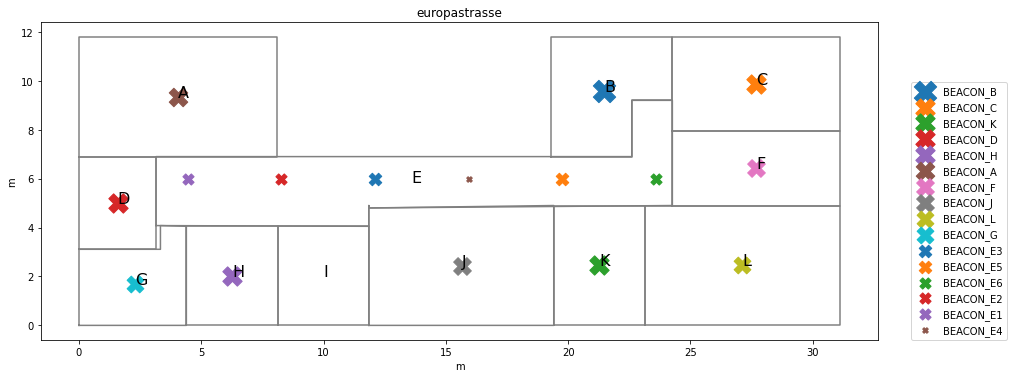}
    \end{subfigure} 
    \caption{Beacon frequency map (Office at Europastrasse): (Left) kNN and (Right) Multilateration}
    \label{fig:beacon_frequency_europa_knn_map}
\end{figure*}

\subsubsection{Model Accuracy vs. Placement of Beacons}
\label{subsub:accuracy-vs-placement}

As a next evaluation step, the interesting question is to analyze the significance of individual beacons concerning accuracy to understand what beacons contribute more to accuracy than others, \ie influence of beacon placement on accuracy. Placing beacons too close or far apart can compromise the system's accuracy. In addition, environmental factors such as interference, signal attenuation, and obstructions can impact the system's accuracy. For an effective indoor tracking system, finding the optimal balance between model accuracy and beacon placement is essential. The objective is to maximize precision while minimizing the number of beacons needed to achieve the desired precision.

Combining this with floor plan information might reveal geospatial patterns for where to place beacons on a floor plan to achieve maximum accuracy. To this end, the frequency is analyzed with which beacons occur in the top-ranking beacon configuration of each feature space length. In this sense, for each group of specific feature-length (\eg all combinations of length 3), the combination is selected which achieves the best accuracy score. 

\begin{figure}[t]
    \centering
    \includegraphics[width=1\columnwidth,trim={0 0 0 0.75cm},clip]{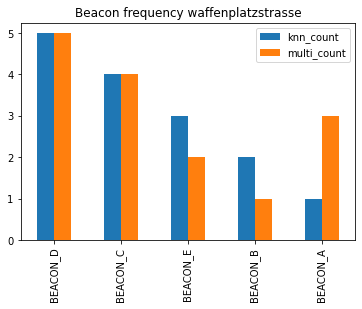}
    \caption{Beacon frequency - Apartment at Waffenplatz}
    \label{fig:beacon_frequency_waffenplatz}
\end{figure}

Repeating this for all feature lengths and counting the occurrence of each beacon yields the bar plots in Figure~\ref{fig:beacon_frequency_europa} and Figure \ref{fig:beacon_frequency_waffenplatz}. 
While the beacon frequency of kNN and multilateralism-based approaches are almost identical in the apartment setting, the office test location differs. For certain beacons, the frequency count for both models is similar and may be off by a count of 1 to 2. However, there are beacons for which the discrepancy in the frequency count is higher. \textit{E.g.}, most beacons in the hallway (E6, E2, E4, and E5) are more important for the fingerprinting-based approach than the multilateration-based one. Figure~\ref{fig:beacon_frequency_europa_knn_map} shows the position of fixed beacons overlaid on the floor plan, where the beacon size corresponds to its frequency count.

It can be seen that beacons D, E6, and F are all aligned horizontally and vertically centered. Further, if beacon A is also considered, two beacons emerge, one on the left (D, A) and one on the right (E6, F). Conversely, Figure \ref{fig:beacon_frequency_europa_knn_map} (right) shows the same map but for the multilateration-based approach, where another pattern emerges. In this case, the beacons in the center of the floor plan (along the hallway) are the least important. Instead, beacons in the outer rooms seem to have higher importance.

The described observations indicate general-pattern candidate locations for each localization approach. In the case of fingerprinting-based localization, beacons might be installed in pairs across a horizontally or vertically centered line concerning the floor plan. Otherwise, beacons might be best placed on the outermost border rooms of the floor plan. Despite those observations, it is important to highlight the qualitative nature of these results. In the case of the apartment, the beacon frequency analysis was not conducted due to the low number of beacons (only 5 beacons).

\subsubsection{Model Accuracy vs. Training Size}
\label{subsub:accuracy-training}

While multilateration is a geometrical approach that does not require prior knowledge of the environment and can work with a relatively small number of reference points, fingerprinting is a statistical method that uses a pre-existing database of signal strength values to determine the location of a device. In this regard, the disadvantage of the fingerprinting method is the need to prepare a data set in which beacons are properly positioned by observing, among other factors, the characteristics of each floor plan and signal reflection. In practice, the theory that the larger (and better, according to the previous points) the training database, the more accurate the model, can be demonstrated.

\begin{figure}[h]
    \centering
    \includegraphics[width=1\columnwidth,keepaspectratio]{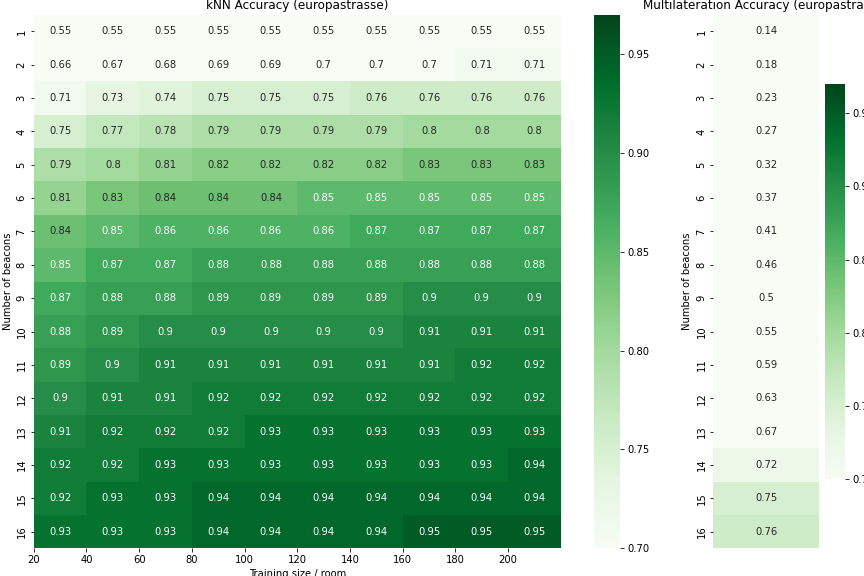}
    \caption{Accuracy scores for the office at Europastrasse}
    \label{fig:accuracy_vs_beacons_europastrasse}
\end{figure}

The following charts present the model accuracy of the kNN localization model subject to the training size. Also, presented results are based on the beacons placement as depicted in Figure \ref{fig:beacon_frequency_europa_knn_map}.
Similar to the previous evaluation, the model was repeatedly trained and tested on different training data sizes, ranging from 20 to 200 samples per room/class. The motivation for this analysis is to reduce the overhead induced by collecting samples for the training database. Figure \ref{fig:accuracy_vs_beacons_europastrasse} depicts a heatmap showing the number of beacons on the y-axis and training size on the x-axis.

As expected, increasing the training size \ie the dataset, positively affects model accuracy, as can be seen by the increasing accuracy score. The kNN model with 16 beacons and a database of 20 records for each beacon results in an accuracy of 93\%. It is important to note that increasing the number of records for each beacon is not as significant in terms of increasing accuracy than adding more BLE beacons, as observed as the number of records increases to a total of 200 records for each beacon, the accuracy reaches a total of 95\%. However, the goal is to achieve a minimum acceptable accuracy with the lowest number of beacons as possible.

\begin{figure}[t]
    \centering
    \includegraphics[width=1\columnwidth,keepaspectratio]{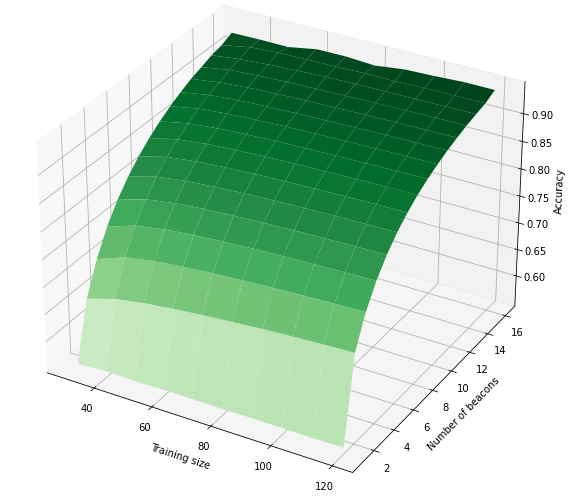}
    \caption{Accuracy curve at the office at Europastrasse}
    \label{fig:accuracy_curve_europastrasse}
\end{figure}

Further to notice is that the positive effect decreases with an increasing number of beacons. For example, the accuracy of a 3-beacon model can be increased by 5\% when using a training set size of 200 instead of 20. In contrast, with a 16-beacon model, the accuracy can only be improved by 2\%. Figure \ref{fig:accuracy_curve_europastrasse} illustrates this behavior in a 3D contour plot. The first derivative of the accuracy vs training size relationship confirms that trend and shows the diminishing marginal effect of additional training samples. In other words, for the configuration utilized in the experiments at the Europastrasse Office, increasing the database has a lesser impact on accuracy than increasing the number of sensors. However, beacon range and/or signal interference can also impact accuracy and necessitate a larger overlap between beacons.

\subsubsection{Economic Analysis}
\label{subsubsec:economic}
With the given analysis on model accuracy, the stage is set for economic considerations regarding the cost of ownership associated with the deployment of each localization algorithm. PMD-Track operates based on mobile phones, which are typically already provided to staff to facilitate coordination and communication, and static and mobile beacons (attached to PMDs). In this sense, the economic analysis only considers the purchase of hardware for the static and mobile tags and activities related to deployment, maintenance, and operation. Table \ref{tab:tag-analysis} compares BLE tag options available on the market and their characteristics.

\begin{table}[b]
\centering
\caption{Comparison of static and mobile beacons}
\label{tab:tag-analysis}
\resizebox{\columnwidth}{!}{%
\begin{tabular}{lllll}
\multicolumn{5}{c}{\textbf{BLE static beacons}}                                                                                                                                                                                                                                                                                                                                                                                                                                                                      \\ \hline
\multicolumn{1}{|l|}{}                                                                 & \multicolumn{1}{l|}{\textbf{\begin{tabular}[c]{@{}l@{}}i10 Durable Beacon\\ \cite{i10Beacon}\end{tabular}}} & \multicolumn{1}{l|}{\textbf{\begin{tabular}[c]{@{}l@{}}E2 Max Beacon\\ \cite{e2Beacon}\end{tabular}}}   & \multicolumn{1}{l|}{\textbf{\begin{tabular}[c]{@{}l@{}}Anchor Beacon\\ \cite{anchorBeacon}\end{tabular}}} & \multicolumn{1}{l|}{\textbf{STiE2 \cite{stie2}}}                                     \\ \hline
\multicolumn{1}{|l|}{\textbf{Vendor}}                                                  & \multicolumn{1}{l|}{Minew, CN}                                                                              & \multicolumn{1}{l|}{Minew, CN}                                                                          & \multicolumn{1}{l|}{Kontakt.io, PL}                                                                       & \multicolumn{1}{l|}{SATO, CN}                                                        \\ \hline
\multicolumn{1}{|l|}{\textbf{Chip}}                                                    & \multicolumn{1}{l|}{\begin{tabular}[c]{@{}l@{}}Nordic nRF52 \\ series\end{tabular}}                         & \multicolumn{1}{l|}{\begin{tabular}[c]{@{}l@{}}Nordic nRF52 \\ series\end{tabular}}                     & \multicolumn{1}{l|}{\begin{tabular}[c]{@{}l@{}}Nordic nRF52 \\ series\end{tabular}}                       & \multicolumn{1}{l|}{\begin{tabular}[c]{@{}l@{}}Nordic nRF52 \\ series\end{tabular}}  \\ \hline
\multicolumn{1}{|l|}{\textbf{Batterylife}}                                             & \multicolumn{1}{l|}{6 years}                                                                                & \multicolumn{1}{l|}{5 years}                                                                            & \multicolumn{1}{l|}{4 years}                                                                              & \multicolumn{1}{l|}{6 years}                                                         \\ \hline
\multicolumn{1}{|l|}{\textbf{Battery type}}                                            & \multicolumn{1}{l|}{AA, replaceable}                                                                        & \multicolumn{1}{l|}{AA, replaceable}                                                                    & \multicolumn{1}{l|}{CR 2477, replaceable}                                                                 & \multicolumn{1}{l|}{AA, replaceable}                                                 \\ \hline
\multicolumn{1}{|l|}{\textbf{Dimensions}}                                              & \multicolumn{1}{l|}{\begin{tabular}[c]{@{}l@{}}73.5 x 44.8 \\ x 24.2 mm\end{tabular}}                       & \multicolumn{1}{l|}{\begin{tabular}[c]{@{}l@{}}72 x 72 \\ x 23 mm\end{tabular}}                         & \multicolumn{1}{l|}{\begin{tabular}[c]{@{}l@{}}56 x 55 \\ x 15 mm\end{tabular}}                           & \multicolumn{1}{l|}{\begin{tabular}[c]{@{}l@{}}71.9 x 71.9 \\ x 24.2mm\end{tabular}} \\ \hline
\multicolumn{1}{|l|}{\textbf{Weight}}                                                  & \multicolumn{1}{l|}{56g}                                                                                    & \multicolumn{1}{l|}{145g}                                                                               & \multicolumn{1}{l|}{35g}                                                                                  & \multicolumn{1}{l|}{50g}                                                             \\ \hline
\multicolumn{1}{|l|}{\textbf{Price}}                                                   & \multicolumn{1}{l|}{< \$12}                                                                                 & \multicolumn{1}{l|}{< \$16}                                                                             & \multicolumn{1}{l|}{< \$36}                                                                               & \multicolumn{1}{l|}{< \$6.50}                                                        \\ \hline
\multicolumn{5}{c}{\textbf{BLE mobile beacons}}                                                                                                                                                                                                                                                                                                                                                                                                                                                                      \\ \hline
\multicolumn{1}{|l|}{}                                                                 & \multicolumn{1}{l|}{\textbf{STiE6   \cite{stie6}}}                                                          & \multicolumn{1}{l|}{\textbf{\begin{tabular}[c]{@{}l@{}}E7 Plus Beacon \\ \cite{e7Beacon}\end{tabular}}} & \multicolumn{1}{l|}{\textbf{H2A \cite{h2abeacon}}}                                                        & \multicolumn{1}{l|}{\textbf{YJ-16019  \cite{HolyIoT}}}                               \\ \hline
\multicolumn{1}{|l|}{\textbf{Vendor}}                                                  & \multicolumn{1}{l|}{SATO,  CN}                                                                              & \multicolumn{1}{l|}{Minew, CN}                                                                          & \multicolumn{1}{l|}{Moko, CN}                                                                             & \multicolumn{1}{l|}{Holyiot, CN}                                                     \\ \hline
\multicolumn{1}{|l|}{\textbf{Chip}}                                                    & \multicolumn{1}{l|}{\begin{tabular}[c]{@{}l@{}}Nordic nRF52 \\ series\end{tabular}}                         & \multicolumn{1}{l|}{\begin{tabular}[c]{@{}l@{}}Nordic nRF52 \\ series\end{tabular}}                     & \multicolumn{1}{l|}{\begin{tabular}[c]{@{}l@{}}Nordic nRF52 \\ series\end{tabular}}                       & \multicolumn{1}{l|}{\begin{tabular}[c]{@{}l@{}}Nordic nRF52 \\ series\end{tabular}}  \\ \hline
\multicolumn{1}{|l|}{\textbf{Battery life}}                                            & \multicolumn{1}{l|}{2 years}                                                                                & \multicolumn{1}{l|}{2 years}                                                                            & \multicolumn{1}{l|}{3 years}                                                                              & \multicolumn{1}{l|}{1-2 years}                                                       \\ \hline
\multicolumn{1}{|l|}{\textbf{\begin{tabular}[c]{@{}l@{}}Battery \\ type\end{tabular}}} & \multicolumn{1}{l|}{\begin{tabular}[c]{@{}l@{}}CR2477, \\ replaceable\end{tabular}}                         & \multicolumn{1}{l|}{\begin{tabular}[c]{@{}l@{}}CR 2477, \\ replaceable\end{tabular}}                    & \multicolumn{1}{l|}{\begin{tabular}[c]{@{}l@{}}CR 2477, \\ replaceable\end{tabular}}                      & \multicolumn{1}{l|}{\begin{tabular}[c]{@{}l@{}}CR 2477, \\ replaceable\end{tabular}} \\ \hline
\multicolumn{1}{|l|}{\textbf{Dimensions}}                                              & \multicolumn{1}{l|}{48 x 48 x 15mm}                                                                         & \multicolumn{1}{l|}{39 x 39 x 15mm}                                                                     & \multicolumn{1}{l|}{48 x 48 x 14mm}                                                                       & \multicolumn{1}{l|}{48 x 48 x 15mm}                                                  \\ \hline
\multicolumn{1}{|l|}{\textbf{Weight}}                                                  & \multicolumn{1}{l|}{23g}                                                                                    & \multicolumn{1}{l|}{20.2g}                                                                              & \multicolumn{1}{l|}{24g}                                                                                  & \multicolumn{1}{l|}{24g}                                                             \\ \hline
\multicolumn{1}{|l|}{\textbf{Price}}                                                   & \multicolumn{1}{l|}{< \$8}                                                                                  & \multicolumn{1}{l|}{<\$5}                                                                               & \multicolumn{1}{l|}{< \$5}                                                                                & \multicolumn{1}{l|}{< \$6}                                                           \\ \hline
\end{tabular}%
}
\end{table}

BLE static beacons do not have the same functionality as gateways or readers, as it is the case with the Impinj xArray device \cite{speedwayconnect} used in the CCount project \cite{RCFSKAS22}. These devices typically have a higher processing power, which allows for the computation of the localization directly on the device, but are more expensive than BLE static beacons (> US\$ 3,000).  In this sense, static beacons are used as reference point for the localization algorithm and work similarly as mobile beacons but with relatively long range and higher battery capacity, as observed in Table \ref{tab:tag-analysis}.

\begin{table}[t]
\centering
\caption{Economical evaluation: parameters}\label{tab:econ_param}
    \begin{tabular}{|l|l|}
    \hline
    \textbf{Parameter} & \textbf{Value} \\ \hline
    Rooms &500 \\  \hline
    Installation time per room &15 min \\ \hline
    Fingerprinting per room &15 min \\ \hline
    Installation hourly rate &\$30 \\ \hline
    Fingerprinting hourly rate &\$30 \\ \hline
    Beacon unit price &\$5 \\ \hline
    Battery unit price &\$2 \\ \hline
    Battery lifetime & 1 year \\ \hline
    beacon-room-factor-knn &0.4 \\ \hline
    beacon-room-factor-multi &0.8 \\ \hline
    \end{tabular}
\end{table}

Table \ref{tab:econ_param} specifies a set of assumptions the following economic evaluation is based. These figures consider the prices charged in Switzerland according to the survey with companies that perform indoor tracking in similar sectors (indoor tracking for marketing analytics, as published in a previous work \cite{RSWTMS22}). First, the scenario envisions the deployment in a location with 500 rooms, corresponding to a typical hospital size. The installation and labeling of a fixed beacon on the ceiling of a room is assumed to be carried out by a trained field worker with an hourly rate of US\$ 30 and an average installation time of 15 minutes. The training time of 15 minutes to collect 200 training samples per room is based on our experience and is also accounted for with an hourly rate of \$30. Hardware components such as beacons and required batteries can be purchased at \$5 and \$2, respectively, at the time of writing. Finally, a beacon-to-room ratio of 0.4 (fingerprinting) and 0.8 (multilateration) was chosen to achieve an estimated accuracy of 80\% and 50-75\%, respectively. These numbers are obtained by extrapolating those observations as of Figure \ref{fig:accuracy_vs_beacons_per_room}.

Table \ref{tab:econ_fingerprinting} compares the associated costs with each localization model. Among the costs associated with setup are the installation of beacons and the collection of fingerprint data. It is important to note that the cost of training is roughly equivalent to the additional beacons required for multilateration. Thus, at first glance, the two approaches appear comparable, with the \textit{multilateration model offering a cost advantage in terms of setup costs being 20\% cheaper than fingerprinting}. However, the annual cost picture changes when maintenance expenses are considered. The annual maintenance costs are driven by the reoccurring cost of batteries and the labor involved in replacing them. In this instance, fingerprinting-based localization has an advantage due to its smaller hardware footprint compared to multilateration, which incurs half the maintenance costs. Considering the total amount with setup and recurring costs, a fingerprinting-based approach is 7\% cheaper than a multilateration-based approach due to the recurring costs. This difference is further increased across multiple years once setup costs are only considered once.

\begin{figure}[b]
    \centering
    \includegraphics[width=1\columnwidth,keepaspectratio]{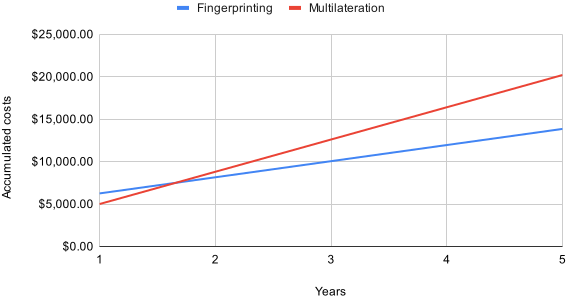}
    \caption{Accumulated costs of ownership}
    \label{fig:costs_of_ownership}
\end{figure}

Figure \ref{fig:costs_of_ownership} outlines the accumulated ownership costs over 5 years. Notable is that the \textit{fingerprinting-based method recovers the higher setup costs in less than two years and has a cost advantage of approximately 45\% cheaper} over five years. Under the current assumptions, it can be concluded that the fingerprinting-based approach is more effective in terms of hardware footprint and cost. This can change once the battery life of beacons is significantly increased, reducing the replacement frequency.

\begin{table}[t]
\centering
\caption{Economical evaluation: comparison between fingerprinting and multilateration}\label{tab:econ_fingerprinting}
\resizebox{\columnwidth}{!}{%
    \begin{tabular}{|l|l|l|l|}
    \multicolumn{4}{c}{Fingerprinting} \\ \hline
    \textbf{Description} & \textbf{Units} & \textbf{Price/Unit} & \textbf{Total} \\ \hline
    BLE Beacons &200 &\$5.00 &\$1,000.00 \\ \hline
    Installation of BLE Beacons &200 &\$7.50 &\$1,500.00 \\ \hline
    kNN Fingerprinting &500 &\$7.50 &\$3,750.00 \\ \hline
    \textbf{Setup Costs} &\textbf{} &\textbf{} &\textbf{\$6,250.00} \\ \hline
    Coin cell battery &200 &\$2.00 &\$400.00 \\ \hline
    Battery replacement work &200 &\$7.50 &\$1,500.00 \\ \hline
    \textbf{Recurring Costs (Yearly)} &\textbf{} &\textbf{} &\textbf{\$1,900.00} \\ \hline
    \multicolumn{4}{c}{Multilateration} \\ \hline
    BLE Beacons &400 &\$5.00 &\$2,000.00 \\ \hline
    Installation of BLE Beacons &400 &\$7.50 &\$3,000.00 \\ \hline
    kNN Fingerprinting &0 &\$7.50 &\$0.00 \\ \hline
    \textbf{Setup Costs} &\textbf{} &\textbf{} &\textbf{\$5,000.00} \\ \hline
    Coin cell battery &400 &\$2.00 &\$800.00 \\ \hline
    Battery replacement work &400 &\$7.50 &\$3,000.00 \\ \hline
    \textbf{Recurring Costs (Yearly)} &\textbf{} &\textbf{} &\textbf{\$3,800.00} \\ \hline
    \end{tabular}
}
\end{table}

\subsubsection{Impacts on Security and Privacy}

\noindent \textit{Security}. Potential attack vectors are the communication link between the BLE beacon and the smartphone, the mobile phone application, and the tracking service itself running at a backend server. Potential impacts are summarized as follows:
\begin{itemize}
    \item A malicious or malfunctioning BLE beacon could be deployed by an adversary to intrude or overload the system in the form of a Denial-of-Service (DoS) attack.
    \item Vulnerability or malfunctioning in the mobile application impairing the device use (DoS) or collecting personal information.
    \item Backend vulnerabilities causing a DoS on the server or emitting incorrect location data.
\end{itemize}

 As mentioned in \cite{boyle2006wireless}, a DoS caused by any device emitting wireless signals can impair the staff communication via radio or their mobile phones via WiFi. BLE supports a variety of communication schemes between two nodes. Some rely on an established connection to exchange encrypted information, whereas others only operate in a broadcast mode, preventing encrypted payloads. Excessive mobile device use due to a vulnerability or programming flaw is also a severe concern, as teams may be unable to communicate promptly. As a result, regardless of the communication scheme used, it is necessary to address these security concerns.

\noindent \textit{Privacy}. Determining where a patient's medical equipment is located raises privacy concerns. For example, tracking the location of a patient's infusion pump also tracks the patient's location. Such concerns must be analyzed on a legal and ethical basis, and mitigation strategies, such as anonymizing patient data, are being developed. Another privacy consideration involves the smartphone and the personnel carrying it. Therefore, the smartphone's location must always be known. This information can be linked to an employee's location, invading his or her privacy. Anonymizing staff identities is a feasible solution to guaranteeing privacy in this situation.

\subsubsection{Limitations and Lessons Learned}
\label{subsub:lessons-and-limitations}

A main limitation of this study was not being able to conduct it within hospital premises with real PMD. The initial intent was to conduct the experimental study in collaboration with the Universitätsspital Zürich USZ; however, it must undergo additional and detailed safety and security analysis before being deployed in an environment with patients, considering that USZ is the major hospital in the Zurich region. As an alternative, a comparison of multilateration and fingerprinting methods for tracking PMDs was made possible by utilizing an office floor with a structure similar to that of multiple hospital rooms.

In addition, PMD-distributed Track's nature presents additional challenges. Regardless of the localization method, the system relies on a critical mass of users to perform distributed PMD tracking. If the number of users is insufficient to cover the entire facility regularly, there may be ``blind zones" where no up-to-date information is accessible due to staff absence. In this case, the deployment of stationary gateways in isolated areas may still be necessary.

In general, the PMD-Track approach demonstrated that it is possible to build a tracking system not relying on stationary gateways or readers as opposed to traditional Real-time Locating Systems (RTLS), while maintaining room-level accuracy. By leveraging smartphones provided to medical staff, it is possible to provide a simpler setup process, lower maintenance overhead, and, most importantly, lower ownership costs compared to existing tracking solutions. Major lessons learned include:
\begin{itemize}
    \item Data pre-processing is the main driver for model accuracy. A straightforward data collection and imputation strategy are necessary to calculate intersections of PMDs tracked, and fixed and mobile beacons (\ie smartphones).
    \item Beacon placement has a significant impact on model accuracy. It is important to consider the types of walls in the room, objects in the room, and possible interference to maximize the RSSI range.
    \item A requirement listed in practice is that tracking approaches that make intensive use of access points, \eg sending probing packets to beacons or mobile phones to check RSSI, are likely not to be used. The WiFi infrastructure is vital to communication within the hospital, and its excessive use may not be permitted by IT staff.
    \item For both test locations, it could be shown that the accuracy of the fingerprinting-based approach surpassed one of the multilateration-based localization algorithms by 15-45\% in non-line-of-sight conditions. Further, it has been shown that fingerprinting-based localization achieves roughly the same level of accuracy with half the number of fixed reference beacons.
    \item Fingerprinting presents 20\% higher costs to setup due to the required training-stage but 50\% lower recurring costs than multilateration as it uses considerably less beacons to achieve the same level of accuracy. A single-year fingerprinting is 7\% cheaper, accumulating to 45\% cheaper than multilateration over the course of five years.
    \item Multilateration is a comparatively simpler approach to deploy and operate than fingerprinting, but has considerably lower tracking accuracy and higher maintenance costs in the experimentation settings used in this paper. 
    \item Hospital IT infrastructures are a critical aspect for the operation of hospitals since various equipment and adequate staff communication. In this sense, indoor tracking approaches should reduce as much as possible the use of access points for RSSI calculation and excessive communication with backend services to avoid a possible DoS of the communication channel.
    \item Simplicity is key. PMD-Track's key strength lies in its lightweight hardware requirements, which yield multiple benefits, such as a simpler setup process, less maintenance overhead, and, most importantly, lower ownership costs than existing tracking solutions.
\end{itemize}

\section{Considerations and Future Work}
\label{sec:conclusions}

PMD-Track provides accurate indoor tracking and inventory management for hospital's infrastructure management. Replacing expensive stationary gateways with BLE beacons and mobile phones provided to the staff provides high accuracy at a relatively low cost than using stationary readers or gateways typically as deployed in traditional tracking approaches. In this regard, PMD-Track evaluated two popular indoor tracking approaches, fingerprinting and multilateration, concerning their accuracy, placement of beacons, and economic impacts. As employees approach tagged PMDs their smartphone updates the location of spotted PMDs in real-time, providing room-level localization data with up to 83\% accuracy for fingerprinting and 35\% for multilateration. The economic analysis yields that fingerprinting presents 7\% cheaper approach in the first year considering setup and recurring costs, which is further increased to a 45\% cheaper than multilateration over five years.

Still, additional fine-tuning and improvements are possible within the proposed PMD-Track approach. Based on the observed results, the fingerprinting-based method was implemented in PMD-Track, and in this sense, although the kNN model worked considerably well with a standard configuration of \(k=7\) and the Euclidean distance metric, different hyperparameters could be evaluated, which might improve the model's accuracy even further. It is also observed that using the users' mobile phones also impacts the battery consumption of these devices, which needs to be evaluated to optimize the application. In this sense, such optimization is considered future work in PMD-Track. Furthermore, an experimental study within a hospital could verify, in practice, if the proposed approach yields greater effectiveness in the daily operations and practical activities of medical staff.


\bibliographystyle{IEEEtranS}
\bibliography{references.bib}
\vspace{-1cm}
\begin{IEEEbiography}[{\includegraphics[width=1in,height=1.25in,clip,keepaspectratio]{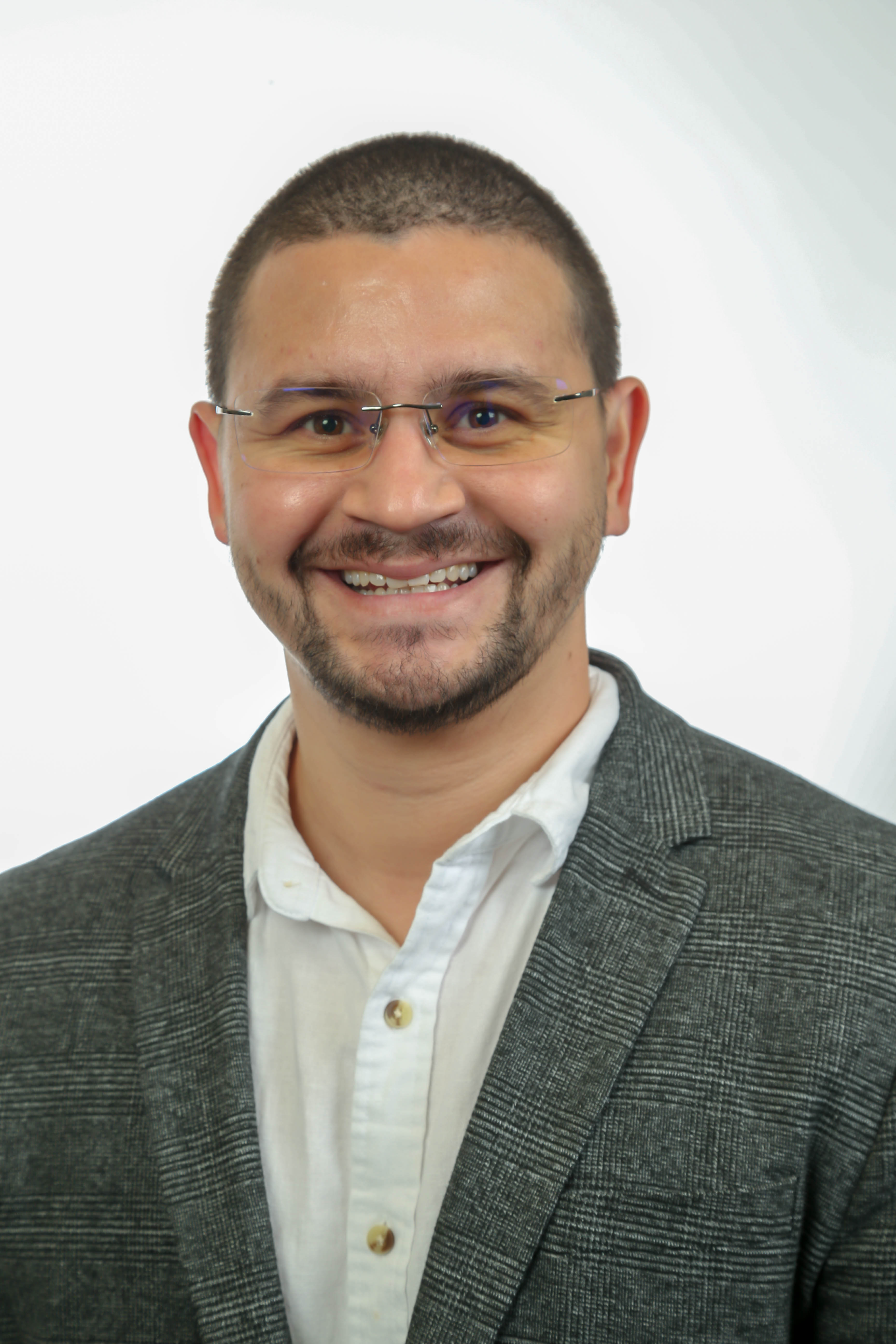}}]{Bruno Rodrigues}
is a Postdoctoral Researcher at University of Z{\"u}rich UZH, Switzerland, within the Communication Systems Group CSG of the Department of Informatics IfI. He received his PhD in 2020 at UZH, focusing on blockchain-based collaborative network defenses, and his MSc from the Polytechnic School of University of São Paulo, Brazil, in 2016, where he worked on research projects in partnership with Ericsson Research focusing on network management based on SDN and energy efficiency. 

Bruno's expertise and research are on collaborative network defenses based on Blockchain, and he works on research projects, such as H2020 CONCORDIA in the scope of cybersecurity and Innossuisse PasWITS in the area of wireless communications and measurements.
\end{IEEEbiography}
\vspace{-1cm}
\begin{IEEEbiography}[{\includegraphics[width=1in,height=1.25in,clip,keepaspectratio]{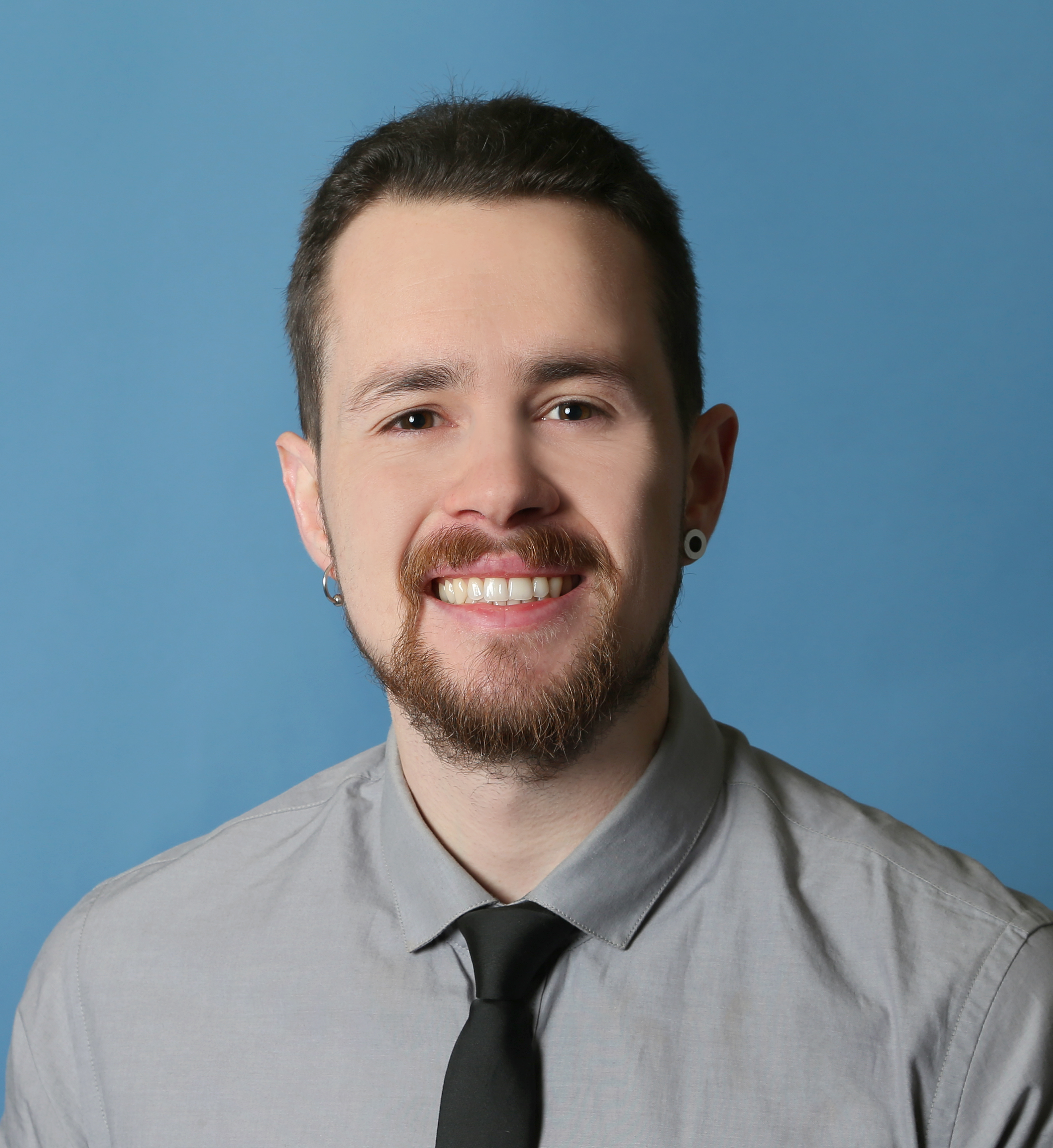}}]{Eder J. Scheid}
received his doctoral degree in 2022 under the supervision of Prof. Dr. Burkhard Stiller at University of Z{\"u}rich UZH, Switzerland, within the Communication Systems Group CSG of the Department of Informatics IfI. Eder holds an MSc degree in Computer Science from the Federal University of the Rio Grande do Sul (UFRGS), Brazil, which he obtained in 2017. 

Eder focuses his research on blockchains, smart contracts, policy- and intent-based network management, and Network Functions Virtualization (NFV).
\end{IEEEbiography}
\vspace{-1cm}
\begin{IEEEbiography}[{\includegraphics[width=1in,height=1.25in,clip,keepaspectratio]{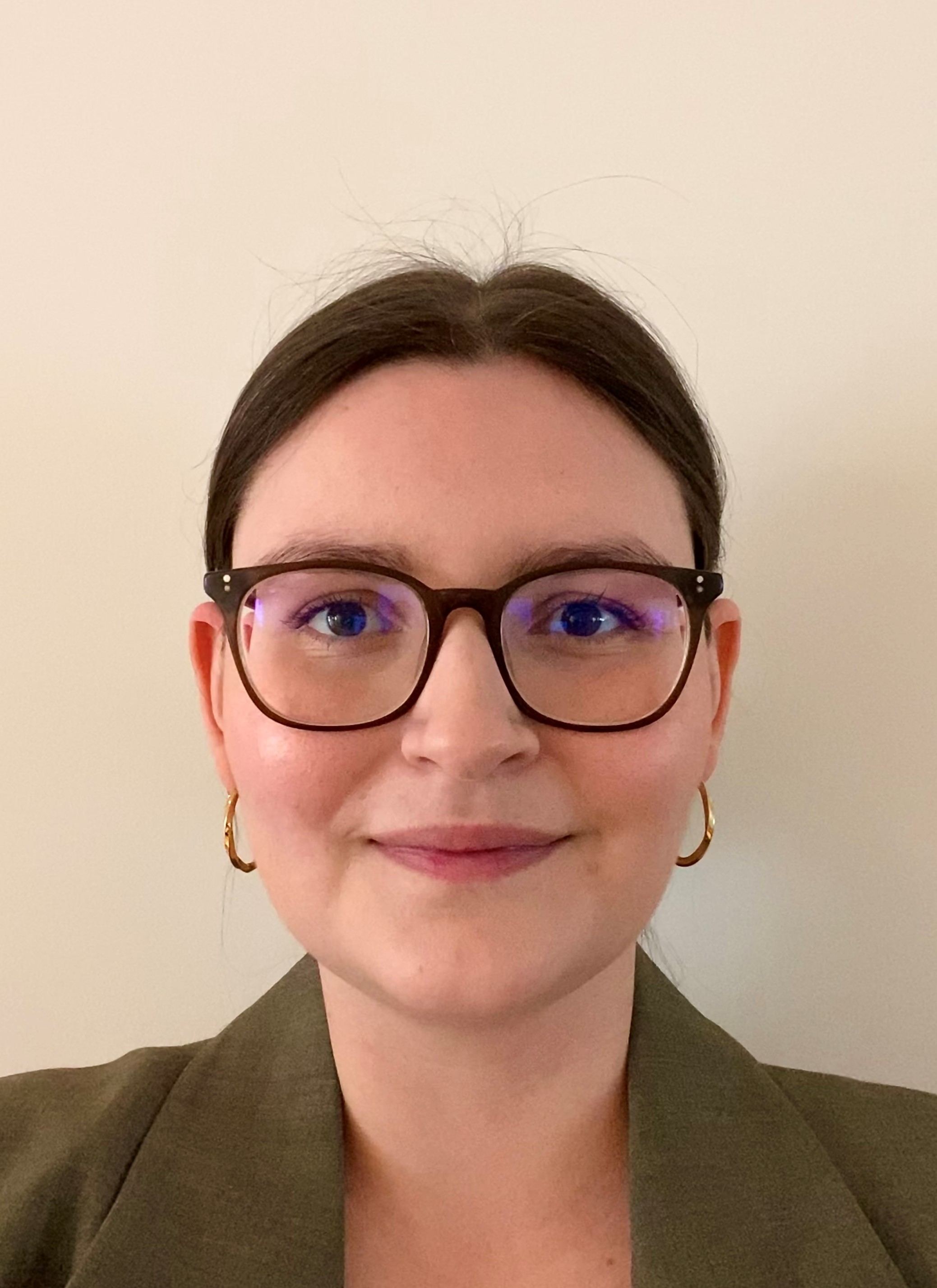}}]{Katharina O. E. Müller}
joined in 2021 the Communication Systems Group CSG at the University of Zürich UZH as a Junior Researcher to pursue her Doctoral Degree under the supervision of Prof. Dr. Burkhard Stiller. Katharina holds an MSc in Computer Science from the Free University Berlin (FU Berlin), Germany, which she obtained in 2021 under the supervision of Prof. Dr. Jochen Schiller. 

Her research interests lie in the Security and Privacy of Internet-of-Things (IoT) devices, standards, protocols, and networks. 
\end{IEEEbiography}
\begin{IEEEbiography}[{\includegraphics[width=1in,height=1.25in,clip,keepaspectratio]{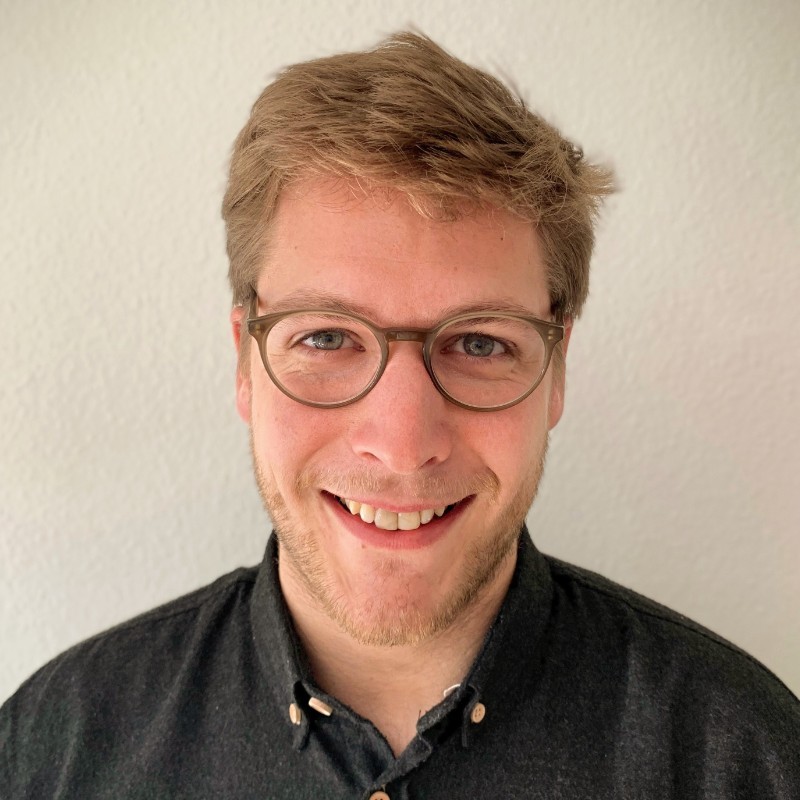}}]{Julius Willems}
is an MSc student in Informatics in the Department of Informatics IfI at the University of Zürich UZH, Switzerland, within
the Communication Systems Group CSG, where he also received his undergraduate BSc degree in Informatics in 2021.
\end{IEEEbiography}
\vspace{-12cm}
\begin{IEEEbiography}[{\includegraphics[width=1in,height=1.25in,clip,keepaspectratio]{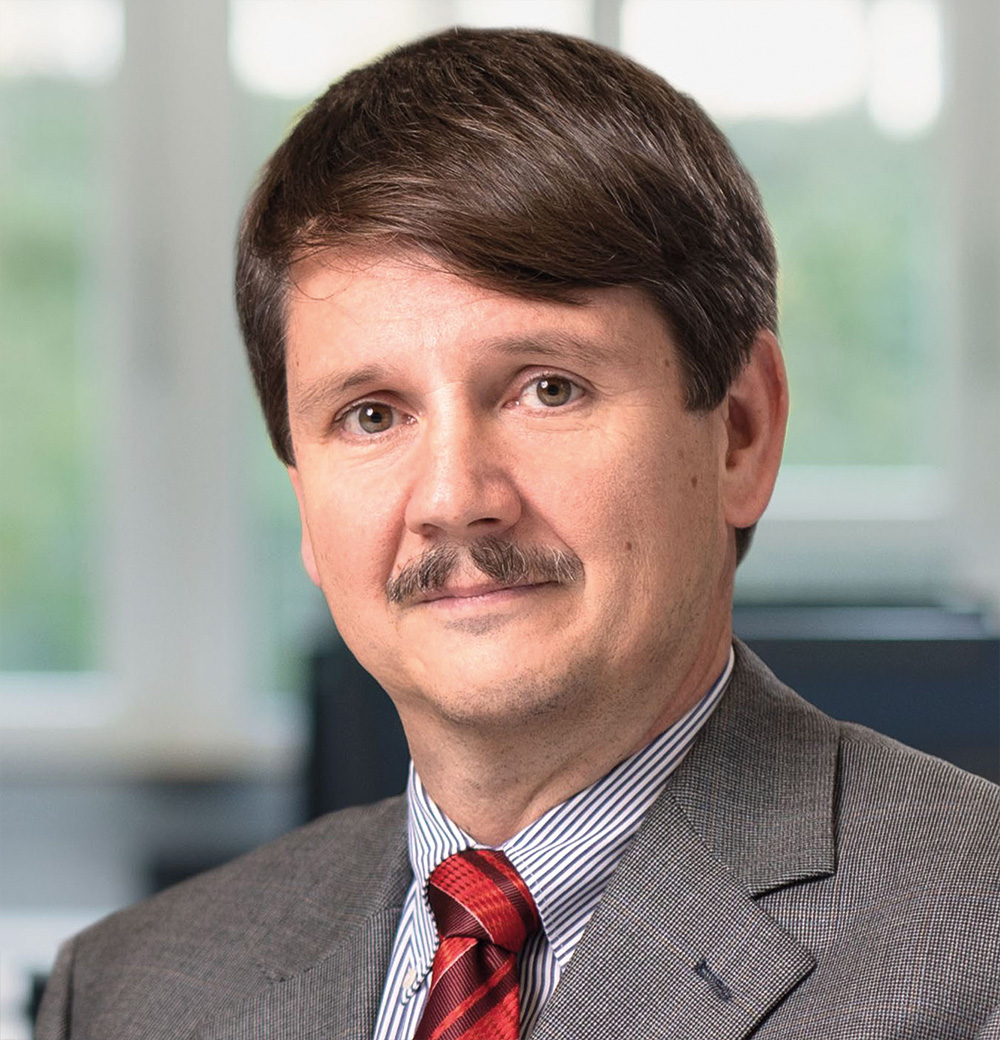}}]{Burkhard Stiller}
received the Informatik-Diplom (MSc) in Computer Science and the Dr. rer.-nat. (PhD) degree from the University of Karlsruhe, Germany, in 1990 and 1994, respectively. In his research career he was with the Computer Lab, University of Cambridge, U.K. (1994-1995), ETH Z{\"u}rich, Switzerland (1995-2004), and the University of Federal Armed Forces Munich, Germany (2002-2004). Since 2004 he chairs the Communication Systems Group CSG, Department of Informatics IfI, at the University of Z{\"u}rich UZH, Switzerland. 

Besides being a member of the editorial board of the IEEE Transactions on Network and Service Management, Springer’s Journal of Network and Systems Management, and the KICS’ Journal of Communications and Networks, Burkhard is the past Editor-in-Chief of Elsevier’s Computer Networks journal, and he runs as the Chair of IFIP's TC6 Committee on Computer Communications. His main research interests are published in well over 300 research papers and include systems with a fully decentralized control (Blockchains, clouds, peer-to-peer), network and service management (economic management), Internet-of-Things (security of constrained devices, LoRa), and telecommunication economics (charging\&accounting).
\end{IEEEbiography}

\EOD
\end{document}